\documentclass[aip,reprint,amsmath,amssymb]{revtex4-2}

\usepackage{graphicx}% figure inclusion
\usepackage[utf8]{inputenc}
\usepackage[T1]{fontenc}
\usepackage{mathptmx}% Times text and math
\usepackage{amsfonts}
\usepackage{xcolor}
\usepackage[colorlinks=true,citecolor=blue,urlcolor=blue,linkcolor=blue]{hyperref}

\usepackage{tikz}
\usetikzlibrary{arrows.meta,calc,bending}

\newcommand{\witnesstikz}{witness_figure_tikz}
\newcommand{\figranks}{rank_window_figure}
\newcommand{\figsweep}{sweep_figure}
\newcommand{\motiftikz}{motif_schematic_tikz}

\begin{document}

\title[``More Is Different'' in Neural Circuits]{Algebraic Emergence of Effective Theories in Canonical Recurrent Motifs of Biological Neuronal Networks}

\author{Nima Dehghani}
\affiliation{McGovern Institute for Brain Research, Massachusetts Institute of Technology, Cambridge, Massachusetts, USA}
\affiliation{The NSF AI Institute for Artificial Intelligence and Fundamental Interactions (IAIFI), Cambridge, Massachusetts, USA}
\email{nima.dehghani@mit.edu}

\date{\today}

\begin{abstract}
Canonical neural circuit motifs are usually described functionally: divisive normalization (DN) rescales population activity by a pooled signal, and winner-take-all competition (WTA) selects one pattern through recurrent excitation and shared inhibition. Here we study these motifs algebraically. We represent them, and their compositions, as finite transformation systems and analyze the transition monoids generated by their input-conditioned updates, distinguishing structure already present in a generator from structure that appears only through composition, and, on a joint state space, structure inherited from one factor from structure that lives on a joint configuration.
Individually aperiodic updates can generate non-aperiodic monoids. In the WTA circuit, every frozen-drive generator collapses all orbits to fixed points, yet short input sequences create local two-cycles of winner-dependent inhibitory gating: globally dissipative dynamics with a reversible action on a small surviving image. The strongest result arises when the selected winner sets the normalization condition. The composed monoid then contains a genuinely composite local cycle in which normalization state and the winner's gating state change together, although every primitive generator is aperiodic. Holonomy analysis certifies this as a group component of the Krohn–Rhodes cascade rather than an incidental cycle, and finds most group-carrying image sets on joint configurations, whereas the uncoupled product has none. An exhaustive interface sweep shows that the composite cycle is a property of the coupling rather than of a chosen map, with exceptions characterized exactly.
Krohn–Rhodes and holonomy methods thus supply both a language for emergent group-like structure and an algebraic contract for composing the motifs. If motifs are building blocks of neural computation, composing them is a form of programming: one chooses primitives and interfaces so that the generated algebra has the intended repertoire. The transition monoid is that repertoire — what a primitive presents to any later construction. Recurrent circuits are compositional transformation systems; their algebra constrains what they can be programmed to compute.\setcounter{mpfootnote}{1}\footnote{For Context \& overview see:\\ \url{https://neurovium.science/posts/pblog-AlgebraicCanonicalNet}}
\end{abstract}

\maketitle

% ----------------------------------------------------------------------
% Introduction
% ----------------------------------------------------------------------
\section{Introduction}

Biological neural circuits compute through the organized evolution of physical states \cite{Dehghani_catcomp_2024,dehghani2026waveCompute}. Recurrent excitation, inhibition, adaptation, gain control, and competition do not just inscribe and modulate representational variables; they also delimit which state transitions are permissible. Constraining which state transitions are possible is in itself a constructor theoretic frame
\cite{Deutsch2013,Deutsch2015} that needs to be addressed if we wish to link structure and function in neural circuits. The mechanisms that stabilize cortical activity are the mechanisms that bound its repertoire — an Andersonian \emph{`More is Different'} \cite{Anderson1972}, now at the scale of a circuit motif. The missing instrument is not another trajectory. It is a language for the computations a motif can generate once its transitions are composed. That question is now operational and is becoming increasingly concrete: connectomes and structure–function maps are used to predict functional organization and to constrain computational models, not merely to catalog wiring \cite{Seung2024WiringDiagram,LappalainenEtAl2024,ShakibaDehghani2026}.

Two such motifs are divisive normalization (DN) and winner-take-all (WTA) competition. Divisive normalization is canonical gain control: a unit’s response is divided by the pooled activity of a surrounding population \cite{Heeger1992,CarandiniHeeger2012}. Introduced for contrast responses in primary visual cortex \cite{Heeger1992}, it recurs in sensory, decision-making, and value computations. LIP (lateral
intraparietal cortex) value signals, for example, are normalized by the values of the available alternatives -- often interpreted as a mechanism for
context-dependent response rescaling \cite{LouieEtAl2011}. Normalization
also interacts with selection-like variables --- the normalization model of
attention makes this explicit \cite{ReynoldsHeeger2009}. Recurrent excitation stabilized by feedback inhibition can itself compute a normalization, as in the stabilized supralinear network \cite{AhmadianRubinMiller2013,RubinVanHooserMiller2015}.

Winner-take-all competition is selection: two excitatory populations, self-excitation, a shared inhibitory pool; one pattern wins, the rest are suppressed \cite{Grossberg1973,Wang2002}. The same excitation–inhibition motif sits at the core of the canonical cortical microcircuit \cite{DouglasMartin2004}, supports coexisting digital selection and analog amplification \cite{Hahnloser2000}, is a computationally powerful primitive in its own right \cite{Maass2000}, and remains stable when WTA modules are coupled \cite{RutishauserDouglasSlotine2011}. 

These two motifs therefore act on neural computation differently.
Normalization changes the effective scale of population responses through gain modulation; competition changes the identity of the selected state. Inhibitory feedback can implement gain control, normalization, and selective amplification inside one recurrent WTA architecture \cite{RutishauserSlotineDouglas2012,RubinVanHooserMiller2015,ReynoldsHeeger2009}. In biological circuits DN and WTA motifs are therefore naturally composable and together they shape the bedrock of contextual state-dependent computation: a normalized field provides an arena for selection, while the selected state can in turn reshape the gain context that follows. Their shared circuitry is not the question. The algebra generated by composing the motifs is.

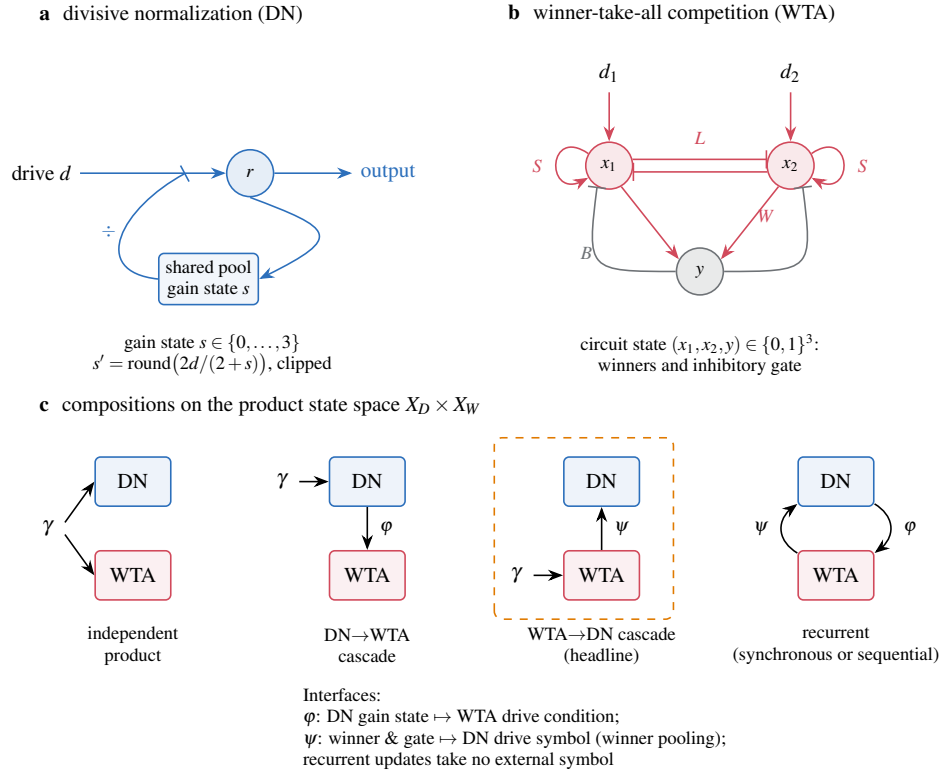
\begin{figure*}[t!]
\centering
% ---------------------------------------------------------------------------
% motif_schematic.tikz -- circuit motifs and their compositions (Fig. 1).
%
% Pure TikZ (not tikz-cd: these are circuit wiring diagrams, not commutative
% diagrams). The including document needs, in its preamble:
%     \usepackage{tikz}
%     \usetikzlibrary{arrows.meta,calc}
% Glyph conventions: filled arrowhead = excitatory / drive connection;
% bar terminal = inhibitory or divisive connection. Colors match the data
% figures (DN blue, WTA red, headline highlight orange).
%
% To render this file alone, compile motif_schematic_standalone.tex, then
%     pdftocairo -svg motif_schematic.pdf motif_schematic.svg
%     pdftocairo -png -r 300 -singlefile motif_schematic.pdf motif_schematic
% ---------------------------------------------------------------------------
\definecolor{dnblue}{HTML}{2C6FBB}
\definecolor{wtared}{HTML}{D1495B}
\definecolor{cycorange}{HTML}{E07A00}
\definecolor{midgrey}{HTML}{6E7276}

\begin{tikzpicture}[
  font=\small,
  exc/.style={-{Stealth[length=2mm]}, semithick},
  inh/.style={-{Bar[width=2.4mm]}, semithick},
  unit/.style={circle, draw, semithick, minimum size=6.2mm, inner sep=0.5pt},
  pool/.style={rectangle, rounded corners=2pt, draw, semithick, inner sep=2.5pt, align=center},
  mod/.style={rectangle, rounded corners=2pt, draw, semithick, minimum width=10mm,
              minimum height=6.5mm, font=\footnotesize},
  dnmod/.style={mod, draw=dnblue, fill=dnblue!8},
  wtamod/.style={mod, draw=wtared, fill=wtared!8},
  lbl/.style={font=\footnotesize},
  tlb/.style={font=\scriptsize},
]

% ===================== (a) divisive normalization ===========================
\node[lbl, anchor=west] at (-1.35,2.55) {\textbf{a}\hspace{5pt}divisive normalization (DN)};
\node[unit, draw=dnblue, fill=dnblue!12] (u)  at (1.55,0.45) {\scriptsize $r$};
\coordinate (din) at (-0.70,0.45);
\node[lbl, anchor=east] at (din) {drive $d$};
\draw[exc, dnblue] (din) -- (u);
\draw[exc, dnblue] (u) -- (2.90,0.45) node[lbl, anchor=west] {output};
\node[pool, draw=dnblue, fill=dnblue!6, tlb] (pl) at (1.00,-0.95) {shared pool\\gain state $s$};
\draw[exc, dnblue] (u.south) .. controls (2.7,-0.30) and (2.8,-.40) .. (pl.east);
\coordinate (syn) at ($(din)!0.62!(u)$);
\draw[inh, dnblue] (pl.west) .. controls (-0.60,-0.85) and (-0.15,0.00) .. (syn);
\node[tlb, dnblue] at (-0.32,-0.32) {$\div$};
\node[tlb, align=center] at (1.05,-1.95) {gain state $s\in\{0,\dots,3\}$\\ $s' = \mathrm{round}\big(2d/(2+s)\big)$, clipped};

% ===================== (b) winner-take-all ==================================
\begin{scope}[shift={(5.9,0)}]
\node[lbl, anchor=west] at (-1.05,2.55) {\textbf{b}\hspace{5pt}winner-take-all competition (WTA)};
\node[unit, draw=wtared, fill=wtared!12] (e1) at (0.40,0.55) {\scriptsize $x_1$};
\node[unit, draw=wtared, fill=wtared!12] (e2) at (2.80,0.55) {\scriptsize $x_2$};
\node[unit, draw=midgrey, fill=midgrey!15] (iu) at (1.60,-0.85) {\scriptsize $y$};
\node[lbl, anchor=south] at (0.40,1.55) {$d_1$};
\draw[exc, wtared] (0.40,1.52) -- (e1.north);
\node[lbl, anchor=south] at (2.80,1.55) {$d_2$};
\draw[exc, wtared] (2.80,1.52) -- (e2.north);
\draw[exc, wtared] (e1.150) .. controls (-0.45,1.10) and (-0.45,0.00) .. (e1.210);
\node[tlb, wtared] at (-0.55,0.55) {$S$};
\draw[exc, wtared] (e2.30) .. controls (3.65,1.10) and (3.65,0.00) .. (e2.330);
\node[tlb, wtared] at (3.75,0.55) {$S$};
\draw[inh, wtared] (e1.15) -- (e2.165);
\draw[inh, wtared] (e2.195) -- (e1.345);
\node[tlb, wtared] at (1.60,0.92) {$L$};
\draw[exc, wtared] (e1.300) -- (iu.150);
\draw[exc, wtared] (e2.240) -- (iu.30);
\node[tlb, wtared] at (2.48,-0.12) {$W$};
\draw[inh, midgrey] (iu.180) .. controls (0.10,-0.85) .. (e1.240);
\draw[inh, midgrey] (iu.0)   .. controls (3.10,-0.85) .. (e2.300);
\node[tlb, midgrey] at (0.10,-0.60) {$B$};
\node[tlb, align=center] at (1.60,-1.95) {circuit state $(x_1,x_2,y)\in\{0,1\}^3$:\\ winners and inhibitory gate};
\end{scope}

% ===================== (c) compositions =====================================
\begin{scope}[shift={(0,-4.25)}]
\node[lbl, anchor=west] at (-1.35,1.62)
  {\textbf{c}\hspace{5pt}compositions on the product state space $X_D\times X_W$};

% --- independent product
\node[dnmod]  (c1d) at (0.00,0.62)  {DN};
\node[wtamod] (c1w) at (0.00,-0.62) {WTA};
\node[lbl, anchor=east] at (-0.95,0.00) {$\gamma$};
\draw[exc] (-0.90,0.08)  -- (c1d.west);
\draw[exc] (-0.90,-0.08) -- (c1w.west);
\node[tlb, align=center] at (0.00,-1.55) {independent\\product};

% --- DN -> WTA cascade
\begin{scope}[shift={(3.1,0)}]
\node[dnmod]  (c2d) at (0,0.62)  {DN};
\node[wtamod] (c2w) at (0,-0.62) {WTA};
\node[lbl, anchor=east] at (-0.95,0.62) {$\gamma$};
\draw[exc] (-0.90,0.62) -- (c2d.west);
\draw[exc] (c2d.south) -- (c2w.north) node[tlb, midway, anchor=west, xshift=1.5pt] {$\varphi$};
\node[tlb, align=center] at (0,-1.55) {DN$\to$WTA\\cascade};
\end{scope}

% --- WTA -> DN cascade (headline)
\begin{scope}[shift={(6.2,0)}]
\draw[dashed, cycorange, semithick, rounded corners=3pt] (-1.42,-1.20) rectangle (0.95,1.20);
\node[dnmod]  (c3d) at (0,0.62)  {DN};
\node[wtamod] (c3w) at (0,-0.62) {WTA};
\node[lbl, anchor=east] at (-0.95,-0.62) {$\gamma$};
\draw[exc] (-0.90,-0.62) -- (c3w.west);
\draw[exc] (c3w.north) -- (c3d.south) node[tlb, midway, anchor=west, xshift=1.5pt] {$\psi$};
\node[tlb, align=center] at (0,-1.55) {WTA$\to$DN cascade\\(headline)};
\end{scope}

% --- recurrent
\begin{scope}[shift={(9.3,0)}]
\node[dnmod]  (c4d) at (0,0.62)  {DN};
\node[wtamod] (c4w) at (0,-0.62) {WTA};
\draw[exc] (c4w.150) .. controls (-0.80,-0.15) and (-0.80,0.15) .. (c4d.210)
    node[tlb, midway, anchor=east, xshift=-1.5pt] {$\psi$};
\draw[exc] (c4d.330) .. controls (0.80,0.15) and (0.80,-0.15) .. (c4w.30)
    node[tlb, midway, anchor=west, xshift=1.5pt] {$\varphi$};
\node[tlb, align=center] at (0,-1.55) {recurrent\\(synchronous or sequential)};
\end{scope}

% \node[tlb, anchor=west] at (-1.35,-2.35)
\node[tlb, align=left] at (5.05,-2.65) 
    {%
      Interfaces:\\
      $\varphi$: DN gain state $\mapsto$ WTA drive condition;\\
      $\psi$: winner \& gate $\mapsto$ DN drive symbol (winner pooling);\\
      recurrent updates take no external symbol%
    };
\end{scope}
\end{tikzpicture}
\caption{
The two canonical motifs and their compositions.
\textbf{(a)} Divisive normalization: a unit driven by input $d$ whose gain is
divisively controlled by a shared pool; the pool's discretized gain state
$s\in\{0,\dots,3\}$ is the DN state variable, updated by
$s'=\mathrm{round}\!\big(2d/(2+s)\big)$ (clipped).
\textbf{(b)} Winner-take-all competition: two excitatory populations
$x_1,x_2$ with self-excitation ($S$), mutual inhibition ($L$), and a shared
inhibitory unit $y$ driven by both ($W$) and feeding back onto both ($B$);
the circuit state is the binary triple $(x_1,x_2,y)$---the two winner bits
and the inhibitory gate.
\textbf{(c)} The four ways the motifs are composed on the product state space:
independent product (shared input $\gamma$, no interaction), the two cascades
(the DN gain state setting the WTA drive through $\varphi$, or the selected
winner and gate setting the DN drive through the winner-pooling map $\psi$),
and full recurrent coupling in which each module's input is set by the other's
state, synchronously or sequentially; in the recurrent schemes the
state-dependent interfaces replace the external symbol, so the
$\gamma$-indexed updates coincide and the update truly takes no external
input. The dashed outline marks the
WTA$\to$DN cascade, the architecture in which composition generates the
paper's central composite group component. Arrowheads denote excitatory/drive
connections; bar terminals denote inhibitory or divisive connections.
}
\label{fig:motifs}
\end{figure*}

The question is whether the motifs generate computational structure(s) beyond their update rules. To make this question
precise, we model each motif's structure-dynamic relation as a deterministic finite transformation system: coarse-grained states acted on by a finite set of input-conditioned transformations. That cut — the link between physical dynamics and effective abstract transitions — is the one structure-preserving accounts of physical computation require \cite{Dehghani_catcomp_2024}. The closure of these map under composition is a transition monoid \cite{EastEgriNagyMitchell2017}:  It contains every transformation that any finite input sequence (\emph{``word''}) can produce. This monoid is an exact algebraic object enabling us to enumerate the full compositional algebra permitted by the circuit. The object of study is therefore the repertoire of the circuit—not one response, one trajectory, or one stimulus condition.

Krohn–Rhodes theory \cite{KrohnRhodes1965,Rhodes2010} is the formal language for analyzing that monoid. Aperiodicity—the absence of nontrivial subgroups—is the classical dividing line \cite{Schutzenberger1965}, and the holonomy form of the decomposition used throughout has a modern self-contained treatment \cite{RhodesSchillingSilva2022}. Every finite transformation semigroup admits, up to division, a decomposition into a cascade of aperiodic reset-like components and finite group components. Under this framework, \emph{aperiodic} parts capture irreversible, memory-erasing transformations. Group components are locally reversible. The appearance of a \emph{nontrivial group component} is therefore not merely a numerical observation about a trajectory; it is a certificate that the finite-state system contains an irreducible local symmetry or cyclic degree of freedom in its transition algebra.

Effective theories of automata already warn that this certificate can be invisible in the microscopic rules: recurrent Boolean circuits can generate higher-level effective structures that are not evident from the microscopic update rules alone \cite{DeDeo2011}. For circuit motifs and their composition, the question is therefore not merely whether the generated monoid becomes larger, but whether qualitatively new ``computational'' structure appears. We make three strict distinctions: \emph{Can aperiodic generators produce a non-aperiodic monoid? Is a local cycle already present in a generator, or only in a word? On a joint state space, does the cycle live in one factor, or on a genuinely joint configuration?}

In neural-computational terms, these ask whether individually dissipative circuit updates can collectively generate a reversible degree of freedom; whether that degree of freedom belongs to one primitive circuit operation or emerges only from a sequence of operations; and whether, once two motifs are coupled, the resulting computation remains attributable to one motif or belongs to the composite circuit itself. The monoid audit and the holonomy decomposition, computed with SgpDec \cite{SgpDec,SgpDecICMS}, are the instruments for making those distinctions exact.

These finite model portraits are probes of the computational repertoire of the cortex, not replicas of cortex. They preserve the logic of gain control, selection, inhibition, and feedback, and they make the compositional closure exactly computable. What a single update contains can then be separated from what appears only under composition. Recurrent circuits are dynamical systems with trajectories. They are also transformation systems, and the algebra of those transformations is part of what they compute. Do canonical circuit motifs such as DN and WTA remain algebraically simple under composition, or can composition generate nontrivial local group structure? Put differently: \emph{is more different in neural circuit motifs?} 

% ----------------------------------------------------------------------
% Results
% ----------------------------------------------------------------------
\section{Results}
\paragraph*{Overview}
We read the three distinctions of the Introduction against two isolated motifs
and then against their compositions. In each case we report both the
transition-monoid audit and the holonomy decomposition computed with SgpDec
\cite{SgpDec,SgpDecICMS}, so that witness-level structure found by exact
enumeration can be matched to its decomposition-level group component in the
Krohn--Rhodes cascade. 

The motifs and the four couplings are drawn in
Fig.~\ref{fig:motifs}. The transition table and functional graph of the
representative WTA generator $f^W_{10}$ are given in
Table~\ref{tab:f10} and Fig.~\ref{fig:f10_sink}, establishing the
fixed-drive sink against which the composed witnesses are read.
The shortest witnesses themselves are shown as functional graphs in
Fig.~\ref{fig:witness}. System-level monoid inventories and holonomy
decompositions are summarized in
Tables~\ref{tab:composition-summary}--\ref{tab:holonomy}, with their rank
structure and the distribution of group-carrying image sets shown in
Fig.~\ref{fig:ranks}. Finally, the exhaustive interface and update-schedule
tests are collected in Fig.~\ref{fig:sweep}.

First, divisive normalization alone. A three-state model already carries an
order-two component in a single generator. That is
\emph{generator-explicit} structure, and precisely the case the rest of the
paper is written to exclude. A four-state model is the stricter object.
Every input-conditioned map is aperiodic; the monoid they generate is not.

Second, winner-take-all: two excitatory populations and a shared inhibitory
pool. Each frozen-drive generator collapses every orbit to a fixed point.
The monoid generated by composing those maps nevertheless contains
non-aperiodic elements. The local group-like action is not a primitive of
the circuit. It is \emph{composition-generated}: produced by switching
among individually dissipative competitive updates.

Third, the two motifs on a joint state space. Four couplings are compared
[Fig.~\ref{fig:motifs}(c)]: the independent product, the DN-to-WTA cascade,
the WTA-to-DN cascade, and recurrent DN--WTA coupling, synchronous and
sequential. The product asks what is inherited by juxtaposition. The two
cascades ask what a directed interface reorganizes. Recurrence asks what
the update schedule itself writes into the generators. The sharp case is
winner-dependent normalization. There the shortest non-aperiodic witness
moves the DN coordinate and the WTA coordinate together --- a
\emph{genuinely composite} cycle --- although every primitive composite
generator remains aperiodic.

\subsection{Single motifs already generate local group structure}

\begin{figure}
    \centering
    \includegraphics[width=\linewidth]{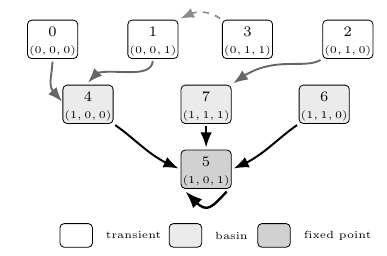}
    \caption{Sink structure of the generator
    $f^W_{10}=(4,4,7,1,5,5,5,5)$ under selective drive to population $x_1$.
    The unique fixed-point attractor is state $5=(1,0,1)$ (dark shading).
    States $4$, $7$, and $6$ form its immediate one-step basin (light shading),
    while $0$, $1$, $2$, and $3$ are transient. The subtle branch is
    $3=(0,1,1)\mapsto 1=(0,0,1)\mapsto 4=(1,0,0)\mapsto 5=(1,0,1)$:
    state $1$ lies in the image of the map, but is not itself an attractor.}
    \label{fig:f10_sink}
\end{figure}

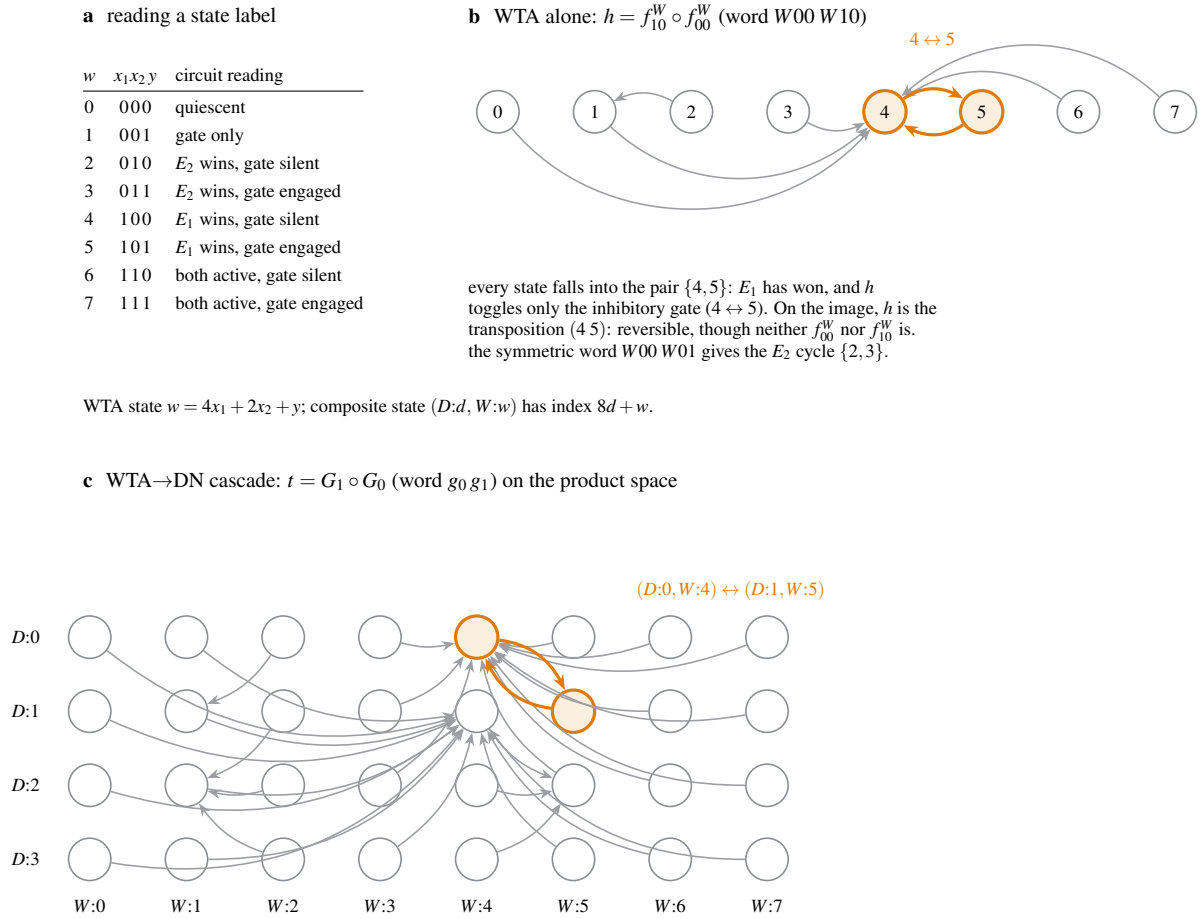
\begin{figure*}[t]
\centering
% ---------------------------------------------------------------------------
% witness_figure_tikz.tex -- GENERATED by gen_witness_figure.py. Do not edit
% arrows by hand: rerun the generator (it verifies them against the audit).
% Needs in the preamble: \usepackage{tikz}
%                        \usetikzlibrary{arrows.meta,calc,bending}
% ---------------------------------------------------------------------------
\definecolor{dnblue}{HTML}{2C6FBB}
\definecolor{wtared}{HTML}{D1495B}
\definecolor{cycorange}{HTML}{E07A00}
\definecolor{midgrey}{HTML}{8A8D91}

\begin{tikzpicture}[
  font=\small,
  st/.style={circle, draw=midgrey, semithick, minimum size=5.6mm, inner sep=0.5pt,
             font=\scriptsize},
  cy/.style={circle, draw=cycorange, very thick, fill=cycorange!12,
             minimum size=5.6mm, inner sep=0.5pt, font=\scriptsize},
  tr/.style={-{Stealth[length=1.7mm]}, midgrey!85, semithick},
  cyc/.style={-{Stealth[length=2mm, bend]}, cycorange, very thick},
  lbl/.style={font=\footnotesize},
  tlb/.style={font=\scriptsize},
]
% ----- panel a: state decoding -------------------------------------
\node[lbl, anchor=west] at (-0.2,2.30)
  {\textbf{a}\hspace{5pt}reading a state label};
\node[anchor=north west, font=\scriptsize] at (-0.2,1.85) {%
  \renewcommand{\arraystretch}{1.12}%
  \begin{tabular}{@{}c@{\hspace{7pt}}c@{\hspace{7pt}}l@{}}
      $w$ & $x_1x_2\,y$ & circuit reading \\[1pt] \hline
      $0$ & $0\,0\,0$ & quiescent \\
      $1$ & $0\,0\,1$ & gate only \\
      $2$ & $0\,1\,0$ & $E_2$ wins, gate silent \\
      $3$ & $0\,1\,1$ & $E_2$ wins, gate engaged \\
      $4$ & $1\,0\,0$ & $E_1$ wins, gate silent \\
      $5$ & $1\,0\,1$ & $E_1$ wins, gate engaged \\
      $6$ & $1\,1\,0$ & both active, gate silent \\
      $7$ & $1\,1\,1$ & both active, gate engaged \\
  \end{tabular}};
\node[tlb, anchor=north west, align=left] at (-0.2,-2.62)
  {WTA state $w=4x_1+2x_2+y$; composite state $(D{:}d,\,W{:}w)$ has index $8d+w$.};
% ----- panel b: WTA witness functional graph -----------------------
\begin{scope}[shift={(5.4,0)}]
\node[lbl, anchor=west] at (-0.5,2.30) {\textbf{b}\hspace{5pt}WTA alone: $h=f^{W}_{10}\circ f^{W}_{00}$ (word $W00\;W10$)};
\node[st] (b0) at (0.00,1.05) {$0$};
\node[st] (b1) at (1.28,1.05) {$1$};
\node[st] (b2) at (2.56,1.05) {$2$};
\node[st] (b3) at (3.84,1.05) {$3$};
\node[cy] (b4) at (5.12,1.05) {$4$};
\node[cy] (b5) at (6.40,1.05) {$5$};
\node[st] (b6) at (7.68,1.05) {$6$};
\node[st] (b7) at (8.96,1.05) {$7$};
\draw[tr] (b0) to[bend right=50] (b4);
\draw[tr] (b1) to[bend right=43] (b4);
\draw[tr] (b2) to[bend right=29] (b1);
\draw[tr] (b3) to[bend right=29] (b4);
\draw[tr] (b6) to[bend right=36] (b4);
\draw[tr] (b7) to[bend right=43] (b4);
\draw[cyc] (b4) to[bend left=35] (b5);
\draw[cyc] (b5) to[bend left=35] (b4);
\node[tlb, cycorange] at (5.75,2.02) {$4\leftrightarrow 5$};
\node[tlb, anchor=north west, align=left] at (-0.5,-1.05)
  {every state falls into the pair $\{4,5\}$: $E_1$ has won, and $h$\\   toggles only the inhibitory gate ($4\leftrightarrow5$). On the image, $h$ is the\\ transposition $(4\;5)$: reversible, though neither $f^{W}_{00}$ nor $f^{W}_{10}$ is.\\   the symmetric word $W00\;W01$ gives the $E_2$ cycle $\{2,3\}$.};
\end{scope}
% ----- panel c: composite witness on the product grid --------------
\begin{scope}[shift={(0,-5.9)}]
\node[lbl, anchor=west] at (-0.2,2.05) {\textbf{c}\hspace{5pt}WTA$\to$DN cascade: $t=G_{1}\circ G_{0}$ (word $g_0\,g_1$) on the product space};
\node[st] (c0) at (0.00,0.00) {};
\node[st] (c1) at (1.28,0.00) {};
\node[st] (c2) at (2.56,0.00) {};
\node[st] (c3) at (3.84,0.00) {};
\node[cy] (c4) at (5.12,0.00) {};
\node[st] (c5) at (6.40,0.00) {};
\node[st] (c6) at (7.68,0.00) {};
\node[st] (c7) at (8.96,0.00) {};
\node[st] (c8) at (0.00,-0.98) {};
\node[st] (c9) at (1.28,-0.98) {};
\node[st] (c10) at (2.56,-0.98) {};
\node[st] (c11) at (3.84,-0.98) {};
\node[st] (c12) at (5.12,-0.98) {};
\node[cy] (c13) at (6.40,-0.98) {};
\node[st] (c14) at (7.68,-0.98) {};
\node[st] (c15) at (8.96,-0.98) {};
\node[st] (c16) at (0.00,-1.96) {};
\node[st] (c17) at (1.28,-1.96) {};
\node[st] (c18) at (2.56,-1.96) {};
\node[st] (c19) at (3.84,-1.96) {};
\node[st] (c20) at (5.12,-1.96) {};
\node[st] (c21) at (6.40,-1.96) {};
\node[st] (c22) at (7.68,-1.96) {};
\node[st] (c23) at (8.96,-1.96) {};
\node[st] (c24) at (0.00,-2.94) {};
\node[st] (c25) at (1.28,-2.94) {};
\node[st] (c26) at (2.56,-2.94) {};
\node[st] (c27) at (3.84,-2.94) {};
\node[st] (c28) at (5.12,-2.94) {};
\node[st] (c29) at (6.40,-2.94) {};
\node[st] (c30) at (7.68,-2.94) {};
\node[st] (c31) at (8.96,-2.94) {};
\node[tlb, anchor=east] at (-0.55,0.00) {$D{:}0$};
\node[tlb, anchor=east] at (-0.55,-0.98) {$D{:}1$};
\node[tlb, anchor=east] at (-0.55,-1.96) {$D{:}2$};
\node[tlb, anchor=east] at (-0.55,-2.94) {$D{:}3$};
\node[tlb, anchor=north] at (0.00,-3.36) {$W{:}0$};
\node[tlb, anchor=north] at (1.28,-3.36) {$W{:}1$};
\node[tlb, anchor=north] at (2.56,-3.36) {$W{:}2$};
\node[tlb, anchor=north] at (3.84,-3.36) {$W{:}3$};
\node[tlb, anchor=north] at (5.12,-3.36) {$W{:}4$};
\node[tlb, anchor=north] at (6.40,-3.36) {$W{:}5$};
\node[tlb, anchor=north] at (7.68,-3.36) {$W{:}6$};
\node[tlb, anchor=north] at (8.96,-3.36) {$W{:}7$};
\draw[tr] (c0) to[bend right=27] (c12);
\draw[tr] (c1) to[bend right=24] (c12);
\draw[tr] (c2) to[bend left=18] (c9);
\draw[tr] (c3) to[bend right=15] (c4);
\draw[tr] (c5) to[bend left=15] (c4);
\draw[tr] (c6) to[bend left=18] (c4);
\draw[tr] (c7) to[bend left=21] (c4);
\draw[tr] (c8) to[bend right=24] (c12);
\draw[tr] (c9) to[bend right=21] (c12);
\draw[tr] (c10) to[bend left=18] (c17);
\draw[tr] (c11) to[bend right=18] (c4);
\draw[tr] (c12) to[bend right=18] (c21);
\draw[tr] (c14) to[bend left=21] (c4);
\draw[tr] (c15) to[bend left=24] (c4);
\draw[tr] (c16) to[bend right=27] (c12);
\draw[tr] (c17) to[bend right=24] (c12);
\draw[tr] (c18) to[bend left=15] (c17);
\draw[tr] (c19) to[bend right=21] (c4);
\draw[tr] (c20) to[bend right=15] (c21);
\draw[tr] (c21) to[bend left=21] (c4);
\draw[tr] (c22) to[bend left=24] (c4);
\draw[tr] (c23) to[bend left=27] (c4);
\draw[tr] (c24) to[bend right=30] (c12);
\draw[tr] (c25) to[bend right=27] (c12);
\draw[tr] (c26) to[bend left=18] (c17);
\draw[tr] (c27) to[bend right=21] (c12);
\draw[tr] (c28) to[bend right=18] (c21);
\draw[tr] (c29) to[bend left=21] (c12);
\draw[tr] (c30) to[bend left=24] (c12);
\draw[tr] (c31) to[bend left=27] (c12);
\draw[cyc] (c4) to[bend left=30] (c13);
\draw[cyc] (c13) to[bend left=30] (c4);
\node[tlb, cycorange, anchor=west] at (7.12,0.62) {$(D{:}0,W{:}4)\leftrightarrow(D{:}1,W{:}5)$};
\end{scope}
\end{tikzpicture}
\caption{
What a two-symbol word does: witness cycles as functional graphs.
\textbf{(a)} How to read a state label. A WTA state $w$ encodes the binary
triple $(x_1,x_2,y)$ --- the two excitatory populations and the inhibitory
gate --- as $w=4x_1+2x_2+y$; a composite state $(D{:}d,\,W{:}w)$ pairs a DN
gain state with a WTA state and has index $8d+w$.
\textbf{(b)} The transformation $h=f^{W}_{10}\circ f^{W}_{00}$ (input word
$W00\,W10$: balanced drive, then drive to population 1) drawn as a functional
graph on the eight WTA states: every arrow is the computed image of a state.
All states are funneled into the pair $\{4,5\}$ --- population 1 has won ---
and on that image $h$ acts as the transposition $(4\;5)$, toggling only the
inhibitory gate. Neither input map alone contains a cycle; the word does. The
symmetric word $W00\,W01$ yields the matching cycle $\{2,3\}$ for population 2.
\textbf{(c)} The composite witness $t=G_{1}\circ G_{0}$ (word $g_0\,g_1$) of
the WTA$\to$DN cascade on the $4\times8$ product grid. The orange two-cycle
$(D{:}0,W{:}4)\leftrightarrow(D{:}1,W{:}5)$ changes the normalization state
and the gate together: the reversible action is supported on a joint circuit
configuration, not on a property of either motif alone. All arrows in (b) and (c) are generated
programmatically from the audited transformations, not drawn by hand.
}
\label{fig:witness}
\end{figure*}

We first ask whether either motif, taken alone, contains reversible structure
that its input-conditioned updates do not. The motifs and the four couplings
are in Fig.~\ref{fig:motifs}. What a single finite-state update \emph{is} is
drawn in Fig.~\ref{fig:f10_sink}: the generator $f^W_{10}$ as a functional
graph, every arrow the image of a state, the whole graph a sink. How a
finite \emph{word} acts is drawn the same way in Fig.~\ref{fig:witness}.
The contrast is the paper's first algebraic fact. One generator has no
cycle of length greater than one; a short word built from two such
generators does. Readers unfamiliar with Krohn--Rhodes theory, holonomy
groups, image sets, or tiles will find a physical primer in
Appendix~\ref{app:krohn_rhodes}; the numerical inventories are in
Tables~\ref{tab:composition-summary} and~\ref{tab:holonomy}.

\paragraph{Three-state DN is the negative control.}
A minimal three-state model of divisive normalization already contains
reversible structure, but the structure is built into a generator
(Methods). Its transition monoid --- every transformation reachable by
composing the input-conditioned updates --- has 13 elements. The group of
units, the globally invertible elements, is isomorphic to $\mathbb{Z}_2$ and
is generated by the intermediate-drive map $f_1=(2,1,0)$, which is its own
inverse. The reversibility is generator-explicit, not composition-generated: $f_1^2=\mathrm{id}$ is
visible without forming any nontrivial word. That is why the three-state
system is the negative control for the rest of the paper. It certifies that
coarse-grained normalization \emph{can} carry an order-two component, and it
fixes the criterion used below: a cycle counts as composition-generated only
when every primitive generator is aperiodic and some element of the generated
monoid is not.

\paragraph{Four-state DN meets the criterion.}
All four input-conditioned maps of the four-state model are aperiodic,
yet the 24-element monoid they generate contains three non-aperiodic
elements --- transformations whose functional graphs contain a nontrivial
cycle. The shortest arises from a three-symbol word and acts as a
transposition on a two-state image. Composition alone creates the structure.
The holonomy decomposition
\cite{Holcombe1982,EgriNagyNehaniv2010} --- the constructive form of the
Krohn--Rhodes cascade, which organizes the monoid's image sets into
hierarchical levels and attaches to each a permutation group acting on its
\emph{tiles}, the maximal image subsets nested within it --- confirms this
at the level of the cascade: the skeleton has depth four and three
components, exactly one of which carries a nontrivial group, a
$\mathbb{Z}_2$ at level three acting on two singleton tiles
(Table~\ref{tab:holonomy}).

\paragraph{WTA generators are sinks; the monoid they generate is not.}
Winner-take-all makes the same point more sharply. The circuit is the
standard selection motif of cortical models --- recurrent self-excitation
of two competing populations checked by a shared inhibitory pool
\cite{Grossberg1973,Hahnloser2000,Wang2002} --- with eight states, spanned
by the two excitatory populations and the inhibitory variable, and four
input-conditioned generators, each of which drives every orbit to a fixed
point. Table~\ref{tab:f10} is the full transition structure of one of
them, $f^W_{10}$, selective drive to population $x_1$;
Fig.~\ref{fig:f10_sink} is the same map drawn as a functional graph.
Under this frozen drive, the generator is purely dissipative: its functional graph is a sink. The unique
fixed point is $5=(1,0,1)$: $x_1$ has won and the inhibitory pool is
engaged. States $4$, $6$, and $7$ lie in its immediate basin. The remaining
configurations are transient and flow into the same attractor; the subtle
branch starts from the state $3$ which does not flow directly to the attractor but instead
passes through the transient state $1$ before entering the
winner basin:
\[
3=(0,1,1)\mapsto 1=(0,0,1)\mapsto 4=(1,0,0)\mapsto 5=(1,0,1),
\]
so even a configuration initially favoring $x_2$ is irreversibly redirected
into the $x_1$ winner state under continued drive $10$. State $1$ belongs
to the image of the $f^W_{10}$ map, but not to the attractor and not to the one-step
basin around it. Repeated iteration of this single generator has no cycle
of length greater than one. 

Their generated monoid has 326 elements,
55 of them idempotent --- transformations satisfying $m^2=m$, so that reapplication produces no further change; equivalently, \(m\) acts as the identity on its own image (if \(y\in\mathrm{Im}(m)\), then \(y=m(x)\), hence $m(y)=m^2(x)=m(x)=y$)---
and its group of units is trivial: there is no nontrivial permutation acting on the full state space. Nevertheless 49 elements are non-aperiodic:
globally irreversible transformations that restrict to a reversible action
on an invariant image. The shortest witness is a two-symbol word ---
balanced input followed by drive to the first population --- whose square
is idempotent with a two-element image, on which the witness acts as a
transposition [Fig.~\ref{fig:witness}(b); Appendix~\ref{app:generators}].
Decoded into circuit variables, the cycle alternates between a state in
which the first population has won with inhibition silent and one in which
it has won with inhibition engaged. The symmetric word produces the
matching cycle for the second population.

\begin{table}[h]
\centering
\setlength{\tabcolsep}{16pt}   % default is 6pt
\begin{tabular}{ccl}
\hline
$q$ & $f^W_{10}(q)$ & State transition \\ 
\hline
0 & 4 & $(0,0,0)\rightarrow(1,0,0)$ \\
1 & 4 & $(0,0,1)\rightarrow(1,0,0)$ \\
2 & 7 & $(0,1,0)\rightarrow(1,1,1)$ \\
3 & 1 & $(0,1,1)\rightarrow(0,0,1)$ \\
4 & 5 & $(1,0,0)\rightarrow(1,0,1)$ \\
5 & 5 & $(1,0,1)\rightarrow(1,0,1)$ (fixed point) \\
6 & 5 & $(1,1,0)\rightarrow(1,0,1)$ \\
7 & 5 & $(1,1,1)\rightarrow(1,0,1)$ \\
\hline
\end{tabular}
\caption{State transitions under input $10$ (drive to $x_1$).}
\label{tab:f10}
\end{table}

\paragraph{The cycle is a gate on a chosen winner, not an exchange of winners.}
The local cycle is therefore not the exchange $(1,0,\cdot)\leftrightarrow
(0,1,\cdot)$ that a symmetry argument would have predicted. It is an
inhibitory gate attached to a winner that has already been chosen. Its
location in the algebra says the same thing: 43 of the 49 non-aperiodic
elements sit at rank two, and none at full rank
[Fig.~\ref{fig:ranks}(a--c); Table~\ref{tab:ranks}]. Local reversibility
appears only after the dynamics have collapsed the state space --- the
circuit first discards the information distinguishing most states, and the
cycle lives in what survives. Selection is irreversible; what it leaves
behind is not.

\paragraph{Holonomy places that cycle on singleton tiles.}
The WTA skeleton has depth seven and fifteen components, two of which
carry a nontrivial group. One, at level five, is a $\mathbb{Z}_2$ on the
image set $\{4,5\}$ whose two tiles are singletons, so it permutes the two
individual states of the witness cycle. The symmetric set $\{2,3\}$ is not
the transversal representative that SgpDec prints but belongs to the same
subduction class \cite{EgriNagyNehaniv2010}, of size fifteen, and carries
an isomorphic $\mathbb{Z}_2$: the two are one algebraic component rather
than two independent ones. The second nontrivial group, at level two, acts
on $\{1,4,5,7\}$ with tiles of size three, and therefore permutes subsets
rather than states --- a distinction we keep explicit throughout, because
only singleton-tiled components license statements about individual circuit
states.

\subsection{Coupling gain control to selection creates a cycle neither motif contains}

Both motifs generate local group structure internally. The question that
motivates the rest of the paper is whether coupling them generates structure
that belongs to neither.

\begin{figure*}[t]
\centering
\includegraphics[width=\textwidth]{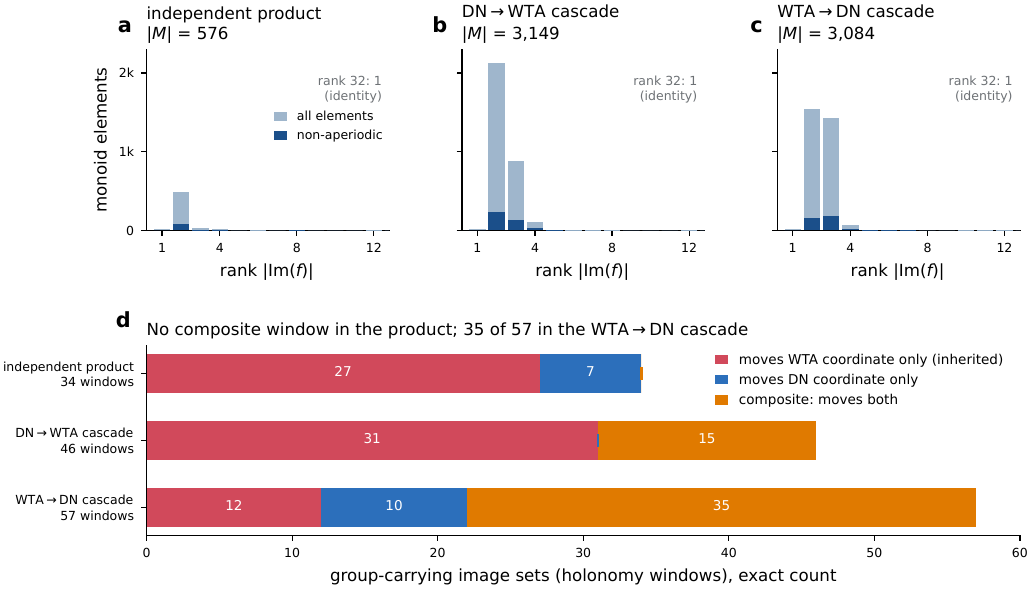}
\caption{
Where the non-aperiodic elements and their local group structure sit in each
composed monoid, from exact enumeration and the holonomy decomposition.
\textbf{(a--c)} Rank distribution $|\mathrm{Im}(f)|$ over every element of the
transition monoid for the independent product ($|M|=576$), the DN$\to$WTA
cascade ($|M|=3149$), and the WTA$\to$DN cascade ($|M|=3084$), with
non-aperiodic elements overlaid (dark). In all three, non-aperiodic elements
concentrate at low rank: local reversible structure appears only on images
reached after irreversible collapse, and only the identity survives at full rank.
\textbf{(d)} The decomposition-level contrast: every image set carrying a
nontrivial permutation group (holonomy window), classified by whether the
group moves only the WTA coordinate (inherited from the WTA factor), only the
DN coordinate, or both (composite). The uncoupled product has $0$ composite
windows of $34$; the DN$\to$WTA cascade has $15$ of $46$; the WTA$\to$DN
cascade is dominated by them, $35$ of $57$. Exact counts from SgpDec
(Table~\ref{tab:holonomy}); panel data are shipped with the code repository.
}
\label{fig:ranks}
\end{figure*}

\paragraph{Four couplings on a $32$-state product.}
We placed the four-state normalization model and the eight-state competition
circuit on a common $32$-state space and compared four ways of joining them:
independent parallel update, normalization driving selection, selection
driving normalization, and full recurrent coupling
[Fig.~\ref{fig:motifs}(c); Table~\ref{tab:composition-summary}]. Rank
inventories and the holonomy classification of group-carrying image sets
are in Fig.~\ref{fig:ranks} and Table~\ref{tab:holonomy}. How to read a
composite state $(D{:}d,\,W{:}w)$ is in Fig.~\ref{fig:witness}(a).

\paragraph{The uncoupled product inherits; it does not create emerging computational composite.}
The independent product is the null model, and it behaves as a null model
should. Its monoid has $576$ elements and $88$ non-aperiodic ones, but every
non-aperiodic witness leaves the normalization coordinate fixed while the
competition coordinate toggles: the structure is the winner-take-all cycle
of the previous section, embedded unchanged in a larger space. Nothing is
created by placing the two motifs side by side. The decomposition makes
this control quantitative. Of the $34$ image sets of the product system
that carry a nontrivial permutation group, $27$ move only the competition
coordinate and $7$ move only the normalization coordinate; \emph{none}
moves both. In the uncoupled product, no group action anywhere in the
algebra sends a state to a state differing in both coordinates.

\begin{table*}[t]
\centering
\begin{ruledtabular}
\begin{tabular}{lcccccc}
System
& States
& Generators
& \(|M|\)
& Idempotents
& Non-ap.
& Shortest witness type \\
\colrule
DN, four-state
& 4 & 4 & 24 & 8 & 3 & DN-local \\
WTA
& 8 & 4 & 326 & 55 & 49 & WTA-local \\
Independent DN--WTA product
& 32 & 4 & 576 & 103 & 88 & WTA-local \\
DN-to-WTA cascade
& 32 & 4 & 3149 & 274 & 401 & WTA-local \\
WTA-to-DN cascade
& 32 & 4 & 3084 & 186 & 361 & genuinely composite \\
Synchronous recurrent DN--WTA
& 32 & 4 (all equal) & 9 & 2 & 6 & generator-explicit composite \\
Asynchronous recurrent, DN first
& 32 & 4 (all equal) & 8 & 2 & 0 & none \\
Asynchronous recurrent, WTA first
& 32 & 4 (all equal) & 8 & 2 & 0 & none \\
\end{tabular}
\end{ruledtabular}
\caption{
Transition-monoid audit of DN, WTA, and DN--WTA composite systems. ``Non-ap.''
denotes the number of non-aperiodic monoid elements. Every system listed has a
trivial group of units; the three-state DN baseline, omitted here, is the sole
exception (\(|M|=13\), units \(\cong\mathbb Z_2\)). In the recurrent systems the
state-dependent interfaces override the external symbol, so the four
$\gamma$-indexed generators coincide and the monoid is cyclic. The strongest
composition-generated result is the WTA-to-DN cascade: every primitive
composite generator is aperiodic, but the generated monoid contains
non-aperiodic elements, and the shortest witness changes both the DN and WTA
coordinates. Explicit generator vectors, state encodings and rank distributions
are given in Appendices~\ref{app:generators} and~\ref{app:ranks}.
}
\label{tab:composition-summary}
\end{table*}

\paragraph{DN-to-WTA enlarges the algebra without changing the shortest witness.}
When the normalization state sets the effective drive to the competition
circuit, the monoid grows more than fivefold, to $3149$ elements with $401$
non-aperiodic ones. Yet the shortest non-aperiodic witness is again
competition-local, with normalization held fixed. Gain control reorganizes
and expands what the circuit can do --- it does not, in this direction,
change the class of the \emph{shortest} non-aperiodic witness. It does make
new structure deeper in the algebra: $15$ of this system's $46$
group-carrying image sets are composite, and three realize the system's single
$\mathbb{Z}_4$ holonomy component--the only group of order greater than two anywhere in our
cascade systems. That $\mathbb{Z}_4$ cycles through
\((D{:}0,W{:}1)\to(D{:}1,W{:}5)\to(D{:}0,W{:}3)\to(D{:}1,W{:}4)\).
Composite structure is present in this direction but not legible: it is
not what a search for the shortest witness returns.

\paragraph{WTA-to-DN makes the shortest witness composite.}
Reversing the coupling does. When the selected winner sets the
normalization condition, all four composite generators remain
irreversible and aperiodic, the monoid has $3084$ elements and $361$ non-aperiodic ones,
and the shortest witness is a two-cycle in which \emph{both} coordinates
change: normalization moves between its two lowest states while the
winner's inhibitory gate opens and closes
[Fig.~\ref{fig:witness}(c); Appendix~\ref{app:generators}]. The winner's
identity is constant throughout; what alternates is its gating state
together with the gain applied to it. This cycle cannot be attributed to
either subsystem. It is not competition-local, because normalization is
not fixed along it, and it is not normalization-local, because the gating
variable moves. It is the first genuinely composite structure that is also
the \emph{shortest} non-aperiodic witness of its system --- the preceding
cascade contains composite group structure, but never at the shortest
witness --- and it is generated entirely by composing maps each of which
is aperiodic.

\paragraph{Holonomy certifies the pair as a group component.}
The pair
\[
C=\{(D{:}0,W{:}4),\,(D{:}1,W{:}5)\}
\]
is an image set of the skeleton at depth $17$ whose permutator group is
$\mathbb{Z}_2$, generated by the transposition exchanging the two states,
and whose holonomy group acts on two singleton tiles --- so the component
permutes these two composite states themselves rather than a coarser
partition. Thirteen elements of the monoid stabilize the pair and exchange
it; one such element maps \((D{:}0,W{:}4)\mapsto(D{:}1,W{:}5)\) and back
while changing both coordinates, which is the sense in which the group
action is composite rather than inherited. Of the $57$ group-carrying
image sets in this system, $35$ are composite, against $12$
competition-local and $10$ normalization-local: composite group structure
is the dominant kind here, whereas in the uncoupled product it does not
occur at all.

\begin{table*}[t]
\centering
\begin{ruledtabular}
\begin{tabular}{lccccl}
System
& Depth
& Components
& Levels with group
& Groups
& Composite windows \\
\colrule
DN, three-state          & 3  & 2  & 1, 2              & $\mathbb Z_2$, $\mathbb Z_2$ & --- \\
DN, four-state           & 4  & 3  & 3                 & $\mathbb Z_2$ & --- \\
WTA                      & 7  & 15 & 2, 5              & $\mathbb Z_2$, $\mathbb Z_2$ & --- \\
Independent product      & 9  & 24 & 2, 7              & $\mathbb Z_2$ (3 components) & 0 of 34 \\
DN-to-WTA cascade        & 12 & 22 & 6, 7, 8, 10       & $\mathbb Z_4$ (1), $\mathbb Z_2$ (4 components) & 15 of 46 \\
WTA-to-DN cascade        & 22 & 92 & 7, 10, 14, 17, 20 & $\mathbb Z_2$ (5 components) & 35 of 57 \\
Synchronous recurrent    & 8  & 7  & 7                 & $\mathbb Z_3$ & --- \\
Asynchronous, DN first   & 9  & 8  & ---               & none & 0 \\
Asynchronous, WTA first  & 9  & 8  & ---               & none & 0 \\
\end{tabular}
\end{ruledtabular}
\caption{
Holonomy decompositions computed with SgpDec. ``Depth'' is the depth of the
skeleton and ``Components'' the number of holonomy components; ``Levels with
group'' lists the levels carrying a nontrivial group component. ``Composite
windows'' counts image sets whose permutation group moves both the
normalization and the competition coordinate, out of all image sets carrying a
nontrivial permutation group; it is the decomposition-level analogue of the
witness classification, and it separates the uncoupled product (none) from the
two cascades. Monoid sizes, idempotent counts, non-aperiodic element counts and
unit-group orders agree with the transition-monoid audit for all nine systems.
}
\label{tab:holonomy}
\end{table*}

\paragraph{The composite group-carrying pair requires two qualifications.}
That pair is
\(C=\{(D{:}0,W{:}4),(D{:}1,W{:}5)\}\), the singleton-tiled
\(\mathbb Z_2\) image set identified above.
First, $C$ is not the transversal representative SgpDec selects for its
subduction class, which has size $45$ and representative
$\{(D{:}1,W{:}2),(D{:}2,W{:}3)\}$; holonomy groups are conjugate along a
class, so $C$ carries the class's $\mathbb{Z}_2$ without appearing in the
component listing. Second, the cascade for this system was not
reconstructed explicitly: its coordinate structure admits
$5.78\times 10^{21}$ tuples and a single cascade element exhausts $14$~GB.
The skeleton, its depth of $22$, all $92$ components, and all five
nontrivial groups are exact, and the decomposition's existence follows
from the holonomy theorem, but we verified the cascade round-trip
elementwise only for the seven smaller systems
(Table~\ref{tab:holonomy}).

\paragraph{Direction, not mere sequence, controls legibility.}
Normalizing and then selecting ($DN \longmapsto WTA$) produces a larger algebra whose composite structure is buried. Selecting and then normalizing according to what was selected ($WTA \longmapsto DN$) produces an algebra in which the composite component is also the most accessible one. Thus, at the biological interfaces studied here, coupling direction does not determine whether composite structure can exist; it determines whether that structure appears as the shortest non-aperiodic witness. Conditioning gain on the outcome of competition brings the composite degree of freedom to the surface.

\subsection{Composite structure is robust across interfaces but shaped by coupling and schedule}

\begin{figure*}[t]
\centering
\includegraphics[width=\textwidth]{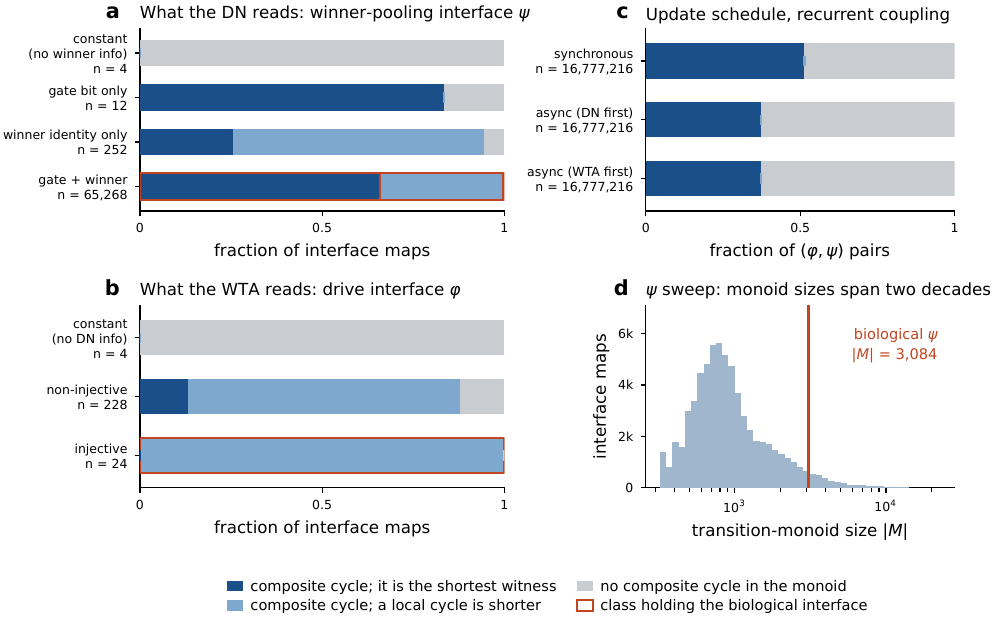}
\caption{
Exhaustive interface and schedule sweeps. \textbf{(a)} Witness class as a
function of what the winner-to-normalization interface $\psi$ reads, over all
$4^8=65\,536$ maps. Constant interfaces carry no winner information and behave
like the uncoupled product. \textbf{(b)} The same classification for the
normalization-to-competition interface $\phi$ over all $4^4=256$ maps: composite
cycles are almost as common, but are almost never the shortest witness, so
legibility tracks coupling direction rather than interface detail.
\textbf{(c)} Update schedule over the full grid of $50\,331\,648$
$(\phi,\psi)$ pairs. \textbf{(d)} Distribution of transition-monoid size across
the $\psi$ sweep; the biological winner-pooling interface sits at the 95.7th
percentile and is not a typical draw. Red outlines mark the class containing the
biological interface.
}
\label{fig:sweep}
\end{figure*}

\paragraph{Two objections, tested by exhaustion.}
Two objections follow immediately: that the effect is an artifact of the
particular map carrying winner identity into normalization, and that it
depends on the update schedule. We tested both by exhaustively enumerating
the interface spaces rather than sampling them
(Fig.~\ref{fig:sweep}).

\paragraph{Composite cycles occur across nearly all WTA-to-DN maps.}
The winner-to-normalization interface $\psi$ assigns one of four drive
symbols to each of eight competition states, giving $4^8=65\,536$ maps; we
enumerated the generated monoid for every one. All $65\,536$ have four
aperiodic generators and a trivial group of units, yet none none of the generated monoids is aperiodic; monoid size ranges from $326$ to $22\,581$ with median
$832$. A composite cycle is present for $65\,278$ of the $65\,536$ maps,
and for $65\,278$ of the $65\,532$ \emph{coupled} maps, a fraction of
$0.9961$ [Fig.~\ref{fig:sweep}(a)].

\paragraph{The exceptions have an exact characterization.}
A composite cycle is absent precisely when the interface never delivers
either extreme drive symbol --- when its image is contained in the two
intermediate symbols --- together with the two constant maps to the
extremes: $258$ maps in total, of which $254$ are coupled. This is an exact
characterization, with no exceptions over the full enumeration. Composite
cycles are therefore not absent only for uncoupled interfaces: $254$
coupled interfaces also lack them, and they are exactly those that cannot
drive normalization to either extreme of its range. All four constant
interfaces, which carry no winner information, behave like the uncoupled
product: competition-local shortest witness, no composite cycle anywhere.

\paragraph{The biological interface is characterized by witness class, not length.}
Its shortest witness is composite, and its length of two is the global
minimum across all $65\,536$ interfaces --- but every interface achieves
length two, so the biological map is not distinguished by witness length
at all. It is distinguished by witness \emph{class}: composite for
$43\,008$ interfaces and competition-local for $22\,528$. That coincidence
holds for $65.6\%$ of coupled interfaces, a majority rather than a
signature [Fig.~\ref{fig:sweep}(a)] --- and for only $64$ of the $252$
nonconstant interfaces that depend only on winner identity, the stratum
matched to the biological winner-pooling interface. The biological
interface is also not a typical draw: its monoid size of $3084$ sits at
the $95.7$th percentile, with only $2816$ interfaces generating a larger
algebra [Fig.~\ref{fig:sweep}(d)].

\paragraph{Direction controls legibility.}
Sweeping the normalization-to-competition interface $\phi$ over its $256$
maps, composite cycles are present for $224$ of the $252$ coupled maps, a
comparable fraction of $0.889$ --- but the shortest witness is composite
for only $30$ of $252$, a fraction of $0.119$ [Fig.~\ref{fig:sweep}(b)].
Presence of composite structure is common in both directions; legibility
differs by more than a factor of five. The biological
normalization-to-competition interface illustrates this: it does contain a
composite cycle, but only at word length three, behind a competition-local
cycle at length two. Every one of the $24$ injective maps in this
direction contains a composite cycle, and not one has it as the shortest
witness.

\paragraph{Synchronous recurrence writes the cycle into the generator.}
The full grid is $256\times65\,536\times3=50\,331\,648$ interface-pair
and schedule combinations [Fig.~\ref{fig:sweep}(c)]. Synchronous recurrent
coupling --- each module setting the other's input at the same instant ---
collapses the algebra to nine elements at the biological interfaces, six
of them non-aperiodic, including a three-cycle spanning both coordinates;
the decomposition assigns it a $\mathbb{Z}_3$, not a $\mathbb{Z}_2$, on
five singleton tiles.

This is a weaker result than it first appears. In the recurrent
constructions the state-dependent interfaces replace the external symbol,
so the four $\gamma$-indexed updates coincide: each recurrent system is
effectively generated by a single transformation. Any non-aperiodicity
must therefore already be present in that recurrent generator rather than
arising through switching among distinct aperiodic generators. The exhaustive
grid confirms this without exception: nowhere among the $50\,331\,648$
recurrent cases does composition generate non-aperiodicity from an
aperiodic recurrent update. Whenever recurrent coupling carries a cycle,
the structure is generator-explicit.

\paragraph{Asynchrony suppresses the structure; it does not abolish it.}
At the biological interfaces both asynchronous variants generate
eight-element monoids with no non-aperiodic element at all, confirmed
aperiodic by the decomposition with no nontrivial group component at
depth nine. Generically, however, asynchrony lowers the composite-cycle
rate from $0.514$ to $0.374$ rather than to zero, leaving $6\,274\,480$
cases per asynchronous scheme with a composite cycle. The two update
orders are numerically indistinguishable in every aggregate we computed.
The complete elimination seen at the biological interfaces is therefore a
property of that operating point, not of asynchronous updating in
general.

\paragraph{The composite degree of freedom occupies a narrow band.}
It is absent in the uncoupled product of motifs. In the WTA-to-DN interface sweep, it is absent when the interface cannot reach the extremes of the
normalization range, apart from the constant extreme controls. Under
recurrent coupling, non-aperiodicity is generator-explicit by construction,
and strict sequential updating suppresses composite cycles substantially.
The cascade therefore occupies a distinct algebraic regime: selection
conditions subsequent normalization while distinct aperiodic input-conditioned
updates remain available for composition. There the composite degree of
freedom is generated rather than inherited or imposed, and in the
winner-to-normalization direction it is also unusually legible.

% ----------------------------------------------------------------------
% Discussion
% ----------------------------------------------------------------------

\section{Discussion}
Canonical recurrent motifs look dissipative under a frozen drive: activity collapses, a winner is selected, gain settles. That is not how they are used and operate in integrated biological circuits. They are driven by changing inputs, recurrent context, and state-dependent feedback \cite{BuonomanoMaass2009}. The computational questions are then three, and they are strict. Can individually dissipative updates collectively generate a reversible degree of freedom? Does that degree of freedom belong to one primitive circuit operation, or does it appear only in a sequence of operations? Once two motifs are coupled, does the resulting computation remain attributable to one motif, or does it belong to the composite circuit itself?

A motif under a frozen input is a map. The same motif under a changing input is a monoid. Representing divisive normalization, winner-take-all competition, and their compositions as finite transformation systems makes those three questions exact: one asks not only what each motif does under a single condition, but which transformations become available under arbitrary finite sequences of inputs. A circuit may have only simple, aperiodic primitive updates, yet the monoid they generate may contain non-aperiodic local structure.

\subsection{From primitive updates to generated computation}

The results separate three levels of structure.  The first is
\emph{generator-explicit} structure, where a nontrivial cycle is already present
in one primitive transformation.  The minimal three-state divisive-normalization
model belongs to this class.  Its intermediate-drive generator acts as an
order-two transformation,
$
f_1^2=\mathrm{id},
$
and the generated monoid contains a global unit isomorphic to
$
\mathbb Z_2.
$
This case is useful because it shows that even a very small normalization motif
can contain reversible algebraic structure after coarse graining.  However, it
does not establish that the group-like component is produced by composition,
because the periodic structure is already present in a generator.

The second level is \emph{composition-generated} structure.  Here every
primitive generator is aperiodic, but some element of the generated monoid is
non-aperiodic:
\[
\forall a,\; f_a \ \text{aperiodic},
\qquad
\exists m\in \langle f_a\rangle
\ \text{non-aperiodic}.
\]
This is the more stringent phenomenon.  It occurs in the four-state
divisive-normalization audit and, more cleanly, in the WTA system.  In the WTA
case, each fixed input condition drives the circuit toward fixed points, yet
switching between input conditions generates a local two-cycle.  Thus the WTA
motif is algebraically richer than any one of its input-conditioned updates.

The third level is \emph{genuinely composite} structure.  This occurs when a
cycle in the composite DN--WTA system cannot be localized to DN alone or WTA
alone, but instead changes both coordinates.  The WTA-to-DN cascade provides the
clearest example:
\[
(D:0,W:4)\leftrightarrow(D:1,W:5).
\]
Here the WTA coordinate toggles between an ungated and gated winner state, while
the divisive-normalization coordinate changes at the same time.  This is the
strongest result in the present analysis because it satisfies both criteria:
the primitive composite generators are aperiodic, and the shortest detected
non-aperiodic witness is genuinely composite. Fig.~\ref{fig:contract} collects these three levels as the algebraic contract of the constructions: a fixed drive selects a primitive map, finite drive sequences generate a monoid, and holonomy locates the provenance and support of any group component of that monoid.

\subsection{The biological meaning of the WTA local cycle}

The local group-like structure found in the WTA system is not a simple exchange
symmetry between the two competing populations.  It is not the cycle
\[
(1,0,\cdot)\leftrightarrow(0,1,\cdot),
\]
which would correspond to alternating winners.  Instead, the shortest WTA
witness [Fig.~\ref{fig:witness}(b)] produces the local cycle
\[
(1,0,0)\leftrightarrow(1,0,1),
\]
or symmetrically,
\[
(0,1,0)\leftrightarrow(0,1,1).
\]
Thus the winner identity is held fixed while the inhibitory or gating variable
toggles.

This matters because it suggests that the relevant algebraic structure of the
WTA motif is not simply selection among alternatives.  Rather, it is selection
together with state-dependent control of the selected state.  Once a winner is
established, the circuit can alternate between a winner-active state and a
winner-gated state.  In biological terms, this resembles the recurrent excitation--inhibition
architecture underlying soft winner-take-all computation, in which a selected
population is maintained while shared inhibitory feedback regulates its
expression
\cite{Hahnloser2000,RutishauserDouglasSlotine2011,RutishauserSlotineDouglas2012}.
The selected population remains the selected population, but its expression is
modulated by inhibitory engagement.

The resulting \(\mathbb Z_2\)-like structure is local rather than global.  The
global group of units of the WTA monoid is trivial, but the monoid contains
low-rank transformations whose images support a two-cycle.  This is precisely
the kind of structure that Krohn--Rhodes and holonomy methods are designed to
detect.  The reversible component is not a global symmetry of the whole state
space; it is a local group action on a subset reached after irreversible
collapse, precisely the kind of structure exposed by holonomy decomposition
\cite{Holcombe1982,EgriNagyNehaniv2010}.

The holonomy decomposition of the WTA monoid confirms this reading.  The
skeleton has depth seven and fifteen components, of which exactly two carry a
nontrivial group, both \(\mathbb Z_2\).  The one at level five acts on the image
set \(\{4,5\}\) --- the support of the witness cycle --- and its two tiles are
singletons, so it permutes those two individual states rather than a coarser
partition of them.  The symmetric set \(\{2,3\}\) belongs to the same subduction
class and carries an isomorphic \(\mathbb Z_2\); the two are one algebraic
component, not two.  The remaining nontrivial group, at level two, acts on
\(\{1,4,5,7\}\) through tiles of size three and therefore permutes subsets, not
states.  We keep that distinction explicit: a holonomy group acts on the tiles
of an image set, and only singleton tiles license a claim about individual
circuit states.

\subsection{Winner-dependent normalization is the strongest compositional case}

Among the DN--WTA composition schemes, the WTA-to-DN cascade is the most
important.  The independent product system contains non-aperiodic elements, but
the shortest witness is WTA-local.  The DN-to-WTA cascade greatly enlarges the
transition monoid, but its shortest witness also remains WTA-local.  In both
cases, the algebraic structure is real, but the clearest cycle can still be
understood as inherited from the WTA subsystem.

The WTA-to-DN cascade is different.  There, the selected WTA state determines
the effective normalization condition.  The shortest witness changes both the
normalization coordinate and the WTA coordinate:
\[
(D:0,W:4)\leftrightarrow(D:1,W:5).
\]
Decoded biologically, this is a transition between a low-normalization,
ungated-winner state and an altered-normalization, gated-winner state.  The
winner remains the same, but the normalization state and inhibitory gating state
co-vary.

This provides a natural interpretation.  Divisive normalization is often understood as a canonical gain-control
operation \cite{Heeger1992,CarandiniHeeger2012}, while WTA circuits implement
competitive selection and selective amplification
\cite{Hahnloser2000,Maass2000,RutishauserDouglasSlotine2011}. Nor are the two
unrelated at the circuit level: canonical-circuit models have shown that
normalization, max-like selection, recurrent excitation, inhibitory pooling,
and modulatory feedback can arise within closely related architectures
\cite{KouhPoggio2008,BroschNeumann2014}. The present result addresses a
different question: what transformation algebra is generated when such
operations are coupled sequentially and state-dependently?  But when WTA
is placed upstream of DN, selection changes the gain-control context itself.
The selected state does not merely emerge from a normalized field; it feeds back
into the normalization process.  Algebraically, this produces a local reversible
component that is not confined to either subsystem.  Biologically, it suggests
that winner-dependent gain control can create an effective computational degree
of freedom not visible in normalization or competition alone.

This is the main compositional claim of the paper.  The result should not be
read as saying that DN and WTA always generate a nontrivial group under any
possible discretization or coupling.  Rather, it shows that under a concrete,
biologically interpretable finite-state abstraction, individually aperiodic
normalization-selection updates can compose into a transition monoid with a
genuinely composite non-aperiodic element.

The holonomy decomposition supports this as the paper's central algebraic
claim.  The pair \(\{(D{:}0,W{:}4),(D{:}1,W{:}5)\}\) is an image set of the
skeleton at depth 17 whose permutator group is \(\mathbb Z_2\) and whose
holonomy group acts on two singleton tiles, so the component exchanges these
two composite states themselves.  Thirteen monoid elements stabilize and
exchange the pair, and an explicit witness changes both coordinates at once.
Across the whole system, 35 of the 57 image sets carrying a nontrivial
permutation group are composite in this sense, against 12 competition-local and
10 normalization-local.  The contrast with the uncoupled product is categorical
rather than quantitative: of the product's 34 group-carrying image sets, not one
moves both coordinates.  Two qualifications belong with the claim.  The pair is
not the transversal representative SgpDec prints for its subduction class, whose
size is 45, so it carries the class's group without appearing in the printed
component list; and the cascade itself was not reconstructed explicitly for this
system, because its coordinate structure admits \(5.78\times10^{21}\) tuples.
The skeleton, its depth, all 92 components and all five groups are exact, and we
verified the cascade round-trip elementwise for the seven smaller systems.

\subsection{Why the product and cascade controls matter}

The independent product and DN-to-WTA systems are important controls.  Without
them, the WTA-to-DN result would be harder to interpret.  The independent
product asks what happens when DN and WTA are merely placed side by side.  Its
non-aperiodic elements show that product systems can inherit local cycles from
their components.  But the shortest witness keeps the DN coordinate fixed, so
the product does not establish genuine compositionality.

The DN-to-WTA cascade asks whether normalization upstream of competition is
sufficient to produce a composite cycle.  This architecture is biologically
natural: divisive normalization can shape the effective field over which
selection and competition act
\cite{ReynoldsHeeger2009,KouhPoggio2008,RubinVanHooserMiller2015}.  Algebraically, this coupling expands the transition monoid substantially,
indicating that DN does reorganize the space of possible transformations.  Yet
the shortest detected cycle remains WTA-local.  This suggests that upstream
normalization enriches the transformation algebra without necessarily producing
the most direct composite group-like witness.

The WTA-to-DN cascade therefore has a special status.  It reverses the causal
direction: the selected winner influences the normalization condition.  This
winner-dependent normalization is what produces the clearest genuinely
composite local cycle.  In this sense, the paper does not simply compare DN and
WTA as separate motifs.  It distinguishes different biological hypotheses about
how normalization and competition are coupled.

\subsection{Recurrent coupling and update schedule}

The recurrent DN--WTA systems show that compositional algebra is sensitive to
update schedule.  In the synchronous recurrent model, composite cycles appear
directly at the level of individual generators.  This means that simultaneous
mutual coupling can insert non-aperiodicity into the primitive update rule
itself.  Such a system is biologically interesting because cortical microcircuits
combine recurrent excitation and inhibitory feedback on overlapping dynamical
timescales, and their responses depend strongly on the state established by
preceding activity
\cite{DouglasMartin2004,RutishauserDouglasSlotine2011,BuonomanoMaass2009}.  However, it is not the
cleanest example of composition-generated algebraic structure, since the
non-aperiodicity is generator-explicit.

By contrast, the asynchronous recurrent variants are fully aperiodic under the
default interface maps.  Both the DN-first and WTA-first asynchronous systems
generate small monoids with no non-aperiodic elements.  This shows that
introducing recurrence alone is not sufficient to generate group-like
structure.  The order in which normalization and competition are updated can
destroy, preserve, or generate local cycles.

More generally, state-dependent selection of subsequent operations is a
central ingredient in formulations of autonomous physical computation
\cite{dehghani2026waveCompute}; here the additional question is what
transformation algebra such state-dependent operations generate under repeated
composition.

This update-schedule dependence is not a weakness of the analysis.  It is part
of the biological message.  Circuit motifs do not have algebraic properties in
isolation from their timing, coupling, and interface maps.  The same component
motifs can yield different transition monoids depending on how they are wired
together.  The finite-transformation framework makes these differences explicit.

The decomposition bears this out and sharpens it.  The synchronous recurrent
system carries a single nontrivial group component, a \(\mathbb Z_3\) rather
than a \(\mathbb Z_2\), supported on five singleton tiles and realizing a
genuine three-cycle on composite states.  Its generators are individually
non-aperiodic, so the group is imposed rather than generated.  Both asynchronous
variants are confirmed aperiodic, with eight components at depth nine and no
nontrivial group anywhere.  An exhaustive sweep of all
\(50\,331\,648\) interface-pair and schedule combinations shows the first of
these findings is not an accident of the biological interface: in no recurrent
case does a non-aperiodic monoid arise from four aperiodic generators.  Under
recurrent coupling, non-aperiodicity is always generator-explicit.  The
asynchronous result is a strong tendency rather than a law --- across the grid,
asynchrony lowers the composite-cycle rate from 0.514 to 0.374 rather than to
zero --- so the complete elimination we observe at the biological interfaces is
a property of that operating point.

\subsection{Relation to Krohn--Rhodes theory}

Krohn--Rhodes theory gives these observations a structural interpretation
\cite{KrohnRhodes1965,Rhodes2010,RhodesSchillingSilva2022}.
A finite transformation semigroup can be decomposed, up to division, into a
cascade of aperiodic components and finite groups.  In this language,
aperiodic components correspond to reset-like or irreversible computation,
whereas group components correspond to locally reversible degrees of freedom
\cite{Schutzenberger1965}.

The systems analyzed here are not large automata constructed for abstract
algebraic purposes.  They are finite-state abstractions of canonical neural
motifs.  The appearance of local group-like structure therefore suggests that
biologically familiar operations---normalization, inhibition, winner selection,
and feedback---can generate the same kinds of algebraic ingredients that appear
in the general structure theory of finite automata.

The distinction between global units and local group components is essential.
In the WTA and DN--WTA examples, the global group of units is trivial.  There is
no permutation symmetry acting on the entire state space.  Nevertheless, the
monoids contain non-aperiodic elements supported on lower-rank images.  A
circuit can therefore be globally dissipative while containing local reversible
components after collapse onto an effective subspace.  This is precisely the
sort of phenomenon that can be missed if one looks only for global symmetries.

The decomposition makes these distinctions concrete.  Every one of the nine
systems has a trivial group of units except the three-state baseline, yet seven
of them contain nontrivial group components: \(\mathbb Z_2\) throughout, with a
single \(\mathbb Z_4\) holonomy component whose action is realized on three
image sets in the DN-to-WTA cascade and a
\(\mathbb Z_3\) in the synchronous recurrent system.  These are group divisors
in the cascade decomposition, not global symmetries, and the three notions must
be kept apart: a non-aperiodic monoid element is a transformation with a
nontrivial cycle; a local permutation group is the action induced on an image
set; and a group divisor is what the decomposition requires as a cascade
factor.  The systems here have the first, the second on low-rank images, and the
third at specific levels of the skeleton, while having no global reversibility at
all.  A further distinction is needed for the group components themselves: they
act on the tiles of an image set, so only where the tiles are singletons does a
component permute individual circuit states.  All the components we interpret
biologically are of that kind.

\subsection{Relation to effective theories of automata}

The analysis connects directly to the idea of effective theories for circuits
and automata \cite{DeDeo2011}.  It also relates to approaches to physical
computation in which abstract computation is defined through a structured
mapping from physical dynamics to abstract transitions, with composition
preserved across levels of description \cite{Dehghani_catcomp_2024}; more
recently, this perspective has been formulated in terms of robust
coarse-grained transition preservation and state-dependent operation selection
\cite{dehghani2026waveCompute}.  The present analysis addresses a
complementary question: once a coarse-grained family of effective
transformations has been specified, what algebra is generated by composing
those transformations?  A
microscopic description of a system may not reveal the effective algebraic
structures that appear after composition, coarse graining, or restriction to
reachable images.  In the present setting, each primitive WTA generator is
aperiodic, but the generated monoid contains local cyclic structure.  Likewise,
the WTA-to-DN cascade contains a composite cycle that is not present in either
coordinate alone as an isolated primitive operation.

This is analogous to the emergence of higher-level effective structure in
finite-state systems.  The local \(\mathbb Z_2\)-like component is not inserted
as a symbolic instruction.  It is produced by the allowed transitions of the
system.  The effective degree of freedom is therefore a property of the
transition algebra, not of a single microscopic rule.

For neural circuits, this provides a useful conceptual bridge. Biological motifs are often described by their immediate functional roles: normalization rescales, WTA selects, inhibition suppresses, recurrence stabilizes. The monoid perspective asks a different question: what computational algebra do these operations generate when composed? Krohn–Rhodes and holonomy analysis supply the language in which that question has an exact answer. The resulting picture is that the composition of neural motifs can produce local group-like structure that is not reducible to a single primitive update. In that sense the paper is a bridge among biological motifs, finite automata, and effective theories of computation: recurrent circuits are not only dynamical systems with trajectories, but transformation systems with a compositional algebra.

\subsection{Interpretation for biological computation}

The biological interpretation should be made carefully.  We are not claiming
that cortical circuits literally instantiate the small finite-state automata
used here.  Nor are we claiming that the detected local cycles correspond
directly to oscillations in continuous neural activity.  The finite systems are
controlled algebraic probes.  They preserve the qualitative logic of canonical
motifs---gain control, competition, inhibition, selection, and feedback---while
making the full transition algebra exactly computable.

Within that controlled setting, the results suggest that canonical neural
motifs can act as generators of effective computational structure.  Divisive
normalization contributes state-dependent gain control
\cite{CarandiniHeeger2012,ReynoldsHeeger2009,ChanceAbbottReyes2002}.  WTA
contributes selection through recurrent excitation and inhibitory feedback
\cite{Hahnloser2000,RutishauserDouglasSlotine2011}.  Their composition can
produce local group-like structure, especially when the selected winner
changes the normalization context.  This gives a formal sense in which
computation can be distributed across motif composition rather than localized
in any single primitive operation.

More broadly, biological structure need not be regarded as an incidental
substrate on which computation is implemented.  Geometry, wiring, and
functional organization can themselves act as computational constraints or
inductive biases \cite{ShakibaDehghani2026}.  The present results make an
analogous point at the level of circuit composition: the organization and
coupling of primitive transformations constrain the computational algebra
available to the circuit.

The WTA-to-DN result is especially suggestive for recurrent cortical circuits.
In recurrent cortical circuits, selected populations need not merely win a
competition and remain passive outputs.  Recurrent excitation and inhibitory
feedback can alter subsequent network responsiveness, while changes in
synaptic context can modulate neuronal gain
\cite{RutishauserSlotineDouglas2012,ChanceAbbottReyes2002,ReynoldsHeeger2009}.  The
finite-state result captures an algebraic analogue of this idea: the selected
state and the normalization state jointly form an effective local degree of
freedom.

\subsection{Programming recurrent motifs}
The same compositional question can be asked in the opposite direction. Circuit
motifs have long served as conceptual and functional building blocks across
levels of neuroscience, linking anatomical organization to computation and
behavior, recurring across neuronal circuit architectures, and even extending
into cellular neurophysiology
\cite{Braganza2018,Luo2021,Mittal2024}.
If canonical motifs are building blocks of neural computation, then composing
them is a form of programming: one chooses primitives and interfaces so that
the algebra generated by their interaction has the intended computational
repertoire. That is a different ambition from fitting a network to a task
after the fact. It requires knowing, before composition, what a primitive can
generate, what structure an interface can preserve or destroy, and what new
structure a coupling can create.

The requirement is sharpest for recurrent systems. Feedforward composition
couples primitives through an interface; in the recurrent constructions
studied here, the interface becomes part of the update rule itself. The
product, cascade, direction, and schedule controls show that the generated
algebra is therefore not a property of the component motifs alone. The same
parts can support inherited, composition-generated, generator-explicit, or
fully aperiodic transformation structure depending on how they are joined and
updated. Without an account of that algebra, the construction of recurrent
motif systems remains largely empirical.

Several complementary developments point toward treating neural systems as
objects that can be constructed compositionally rather than only trained and
interrogated afterward. Neural-engineering frameworks begin from specified
representations, transformations, and dynamics and derive neural
implementations \cite{EliasmithAnderson2003}. Reusable dynamical motifs have
been identified as units of compositional computation in recurrent networks
\cite{Driscoll2024}, while neural machine-code approaches demonstrate that
desired computations can be compiled directly into recurrent substrates
\cite{Kim2023}. From the formal side, the expressivity of recurrent models can
itself be posed algebraically, with computational capability characterized
through syntactic monoids and wreath-product structure
\cite{Nowak2026}. These approaches address different levels of the same
constructive problem.

The missing counterpart for circuit composition is an algebraic description
of the primitives themselves. At the finite-state level, the transition
monoid of a motif --- and of a particular motif coupling --- provides such a
description: an exact account of the transformation repertoire that the
building block presents to any subsequent composition. 

A programming layer
for recurrent neural systems therefore needs both a means of specifying or
compiling a desired computation and an \emph{algebraic contract} for the
pieces being composed [Fig.~\ref{fig:contract}]. The former states what the system should compute; the
latter states what its primitives and interfaces can actually generate.

\subsection{From connectomes to transformation repertoires}

Dense connectomes have changed what it is possible to ask about circuit motifs.
Classical motif analysis established that local cortical connectivity is
highly nonrandom \cite{SongEtAl2005} and distinguished structural motifs from
the repertoire of functional motifs a given wiring pattern can support
\cite{SpornsKotter2004}. The scale of that structural inventory is now
brain-wide: the adult \emph{Drosophila} connectome comprises roughly $140{,}000$
neurons and tens of millions of chemical synapses
\cite{DorkenwaldEtAl2024}, with cell types annotated across the whole brain
\cite{SchlegelEtAl2024}; in mouse visual cortex, functional recordings from
some $75{,}000$ neurons have been co-registered with an electron-microscopy
reconstruction exceeding $200{,}000$ cells \cite{MICrONS2025}; and a petavoxel
fragment of human cortex has been reconstructed at nanoscale resolution
\cite{ShapsonCoeEtAl2024}. Yet the central challenge is not merely to catalogue connectivity, but to
infer what computations that connectivity can support. Recent work has begun
to interpret neuronal wiring diagrams directly as sources of functional
hypotheses \cite{Seung2024WiringDiagram}, to construct dynamical models
constrained by measured connectivity \cite{LappalainenEtAl2024}, and to use
cortical geometry, anatomical connectivity, and function measured in the same
tissue as inductive biases for recurrent neural networks
\cite{ShakibaDehghani2026}. These approaches move from structure to functional
prediction, and from structure and function to constrained dynamics or
learning; the complementary question posed here is what computational
repertoire is generated once a circuit and its effective transformations have
been specified.

The formalism developed here supplies a complementary missing layer, at the
level of sequential computation rather than dynamics. Once a biologically
justified coarse-graining and an input-conditioned update family are assigned
to a connectomic subcircuit --- a candidate normalization pool, a competitive
triad, a coupled pair --- exact monoid enumeration and holonomy analysis
classify its computational repertoire by invariants: aperiodicity, rank
structure, local group components, and whether those components are inherited
from a subcircuit or generated by the coupling. The pipeline is
\begin{align*}
\text{connectome}
  &\rightarrow \text{candidate circuit} \nonumber\\
  &\rightarrow \text{coarse-grained transformations} \nonumber\\
  &\rightarrow \text{generated monoid},
\end{align*}
and the present paper analyzes its final stage for two widely studied
canonical circuit motifs. We emphasize what this does not claim: no
part of the present work derives the update maps from a measured connectome,
and the interface sweep shows that the algebraic repertoire depends on
coupling structure that anatomy alone underdetermines. What the formalism
offers is a precise target for that missing information --- two circuits with
identical wiring diagrams but different effective interfaces can generate
algebras as different as an aperiodic monoid and one carrying composite group
components. The present results extend this perspective to the level of
circuit composition, moving from the structural phenotype provided by a
connectome toward a formal characterization of the circuit's computational
repertoire.

\begin{figure*}[t]
\centering
\resizebox{\textwidth}{!}{% Algebraic contract of the constructions (Discussion / Conclusion).
% for \input from a figure* in main_final.tex.
\begin{tikzpicture}[
  font=\footnotesize,
  >=Stealth,
  every node/.style={align=center},
  pipe/.style={
    draw=black!55,
    fill=black!4,
    rounded corners=1.5pt,
    inner sep=3.4pt,
    text width=3.58cm,
    minimum height=0.92cm
  },
  qbox/.style={
    draw=black!70,
    fill=white,
    rounded corners=1.3pt,
    inner sep=3.2pt,
    text width=2.46cm,
    minimum height=1.08cm
  },
  leaf/.style={
    draw,
    rounded corners=1.3pt,
    inner sep=3.0pt,
    text width=5.15cm,
    minimum height=1.18cm
  },
  genex/.style={leaf, draw=blue!50!black, fill=blue!7},
  local/.style={leaf, draw=magenta!65!black, fill=magenta!6},
  buried/.style={leaf, draw=black!55, fill=black!5},
  comp/.style={leaf, draw=orange!75!black, fill=orange!16},
  aper/.style={leaf, draw=black!45, fill=black!6},
  arr/.style={-Stealth, line width=0.55pt, draw=black!60},
  lab/.style={font=\scriptsize, text=black!52}
]

% ------------------------------------------------------------------
% (a) pipeline
% ------------------------------------------------------------------
\node[pipe] at (0.00,0) (drive)
  {fixed drive\\[-0.6pt]
   {\scriptsize primitive map $f_a:X\to X$}};
\node[pipe] at (4.18,0) (monoid)
  {drive sequences\\[-0.6pt]
   {\scriptsize$M=\langle f_a:a\in\Sigma\rangle$}};
\node[pipe] at (8.36,0) (hol)
  {holonomy decomposition\\[-0.6pt]
   {\scriptsize group components and tiles}};
\node[pipe] at (12.54,0) (read)
  {algebraic contract\\[-0.6pt]
   {\scriptsize provenance and support}};

\draw[arr] (drive) -- (monoid);
\draw[arr] (monoid) -- (hol);
\draw[arr] (hol) -- (read);

\node[lab, anchor=south west] at (-1.88,0.62) {\textbf{(a)} algebraic contract};

% ------------------------------------------------------------------
% (b) three distinctions
% ------------------------------------------------------------------
\node[lab, anchor=south west] at (-1.88,-1.12) {\textbf{(b)} the three distinctions};

\node[qbox] at (0.00,-2.22) (q1) {are all generators\\aperiodic?};
\node[qbox] at (0.00,-4.95) (q2) {is $M$ aperiodic?};
\node[qbox] at (0.00,-8.62) (q3) {where does the\\shortest cycle live?};

\node[genex] at (5.48,-1.58) (dn3)
  {\textbf{generator-explicit}\\[-0.5pt]
   DN, three-state: $f_1^2=\mathrm{id}$
   \ {\scriptsize(units $\cong\mathbb{Z}_2$)}};
\node[genex] at (5.48,-2.96) (sync)
  {synchronous recurrent DN--WTA\\[-0.5pt]
   the cycle is already in the map
   \ {\scriptsize(holonomy $\mathbb{Z}_3$)}};

\node[aper] at (5.48,-4.95) (async)
  {\textbf{monoid aperiodic}\\[-0.5pt]
   asynchronous recurrent\\[-0.4pt]
   {\scriptsize biological interfaces: no group component}};

\node[local] at (5.48,-6.85) (wta)
  {\textbf{local / factor-local shortest witness}\\[-0.5pt]
   DN$_4$;\ WTA gate $\{4,5\}$;\\[-0.4pt]
   uncoupled product (inherited)\\[-0.4pt]
   {\scriptsize product: $0$ of $34$ windows composite}};
\node[buried] at (5.48,-8.62) (d2w)
  {\textbf{factor-local shortest;}\\[-0.6pt]
   \textbf{composite deeper}\\[-0.5pt]
   DN$\to$WTA at biological $\varphi$\\[-0.4pt]
   {\scriptsize one $\mathbb{Z}_4$ component;\ $15$ of $46$ windows composite}};
\node[comp] at (5.48,-10.48) (w2d)
  {\textbf{genuinely composite shortest witness}\\[-0.5pt]
   WTA$\to$DN at biological $\psi$:\\[-0.4pt]
   $(D{:}0,W{:}4)\leftrightarrow(D{:}1,W{:}5)$\\[-0.4pt]
   {\scriptsize $\mathbb{Z}_2$ on two singletons;\ $35$ of $57$ composite}};

% tree edges
\draw[arr] (q1.south) -- (q2.north)
  node[midway, lab, left=1.5pt] {yes};
\draw[arr] (q2.south) -- (q3.north)
  node[midway, lab, left=1.5pt] {no};

\draw[arr] (q1.east) -- ++(0.42,0) |- (dn3.west)
  node[pos=0.1, lab, above=0.5pt, left=-2pt] {no};
\draw[arr] (q1.east) -- ++(0.42,0) |- (sync.west);

\draw[arr] (q2.east) -- (async.west)
  node[midway, lab, above=0.5pt] {yes};

\coordinate (q3j) at ($(q3.east)+(0.42,0)$);
\draw[arr] (q3.east) -- (q3j) |- (wta.west);
\draw[arr] (q3.east) -- (q3j) |- (d2w.west);
\draw[arr] (q3.east) -- (q3j) |- (w2d.west);

% programming note
\node[
  draw=black!40,
  dashed,
  rounded corners=1.5pt,
  inner sep=5.2pt,
  text width=3.55cm,
  anchor=north west,
  align=left
] at (8.20,-1.18) (prog)
  {\scriptsize
   Programming the motifs means choosing primitives and an interface so that the composite lands on the intended leaf of (b).\\[3.2pt]
   Juxtaposition (the uncoupled product) \emph{inherits}.\\[1.6pt]
   Winner-dependent normalization (WTA$\to$DN) \emph{generates}.\\[3.2pt]
   Direction controls legibility, not existence: at the biological interfaces, DN$\to$WTA already contains composite group structure, but its shortest witness remains WTA-local.};

\end{tikzpicture}}
\caption{
The algebraic contract of the constructions.
\textbf{(a)} A fixed drive selects a primitive input-conditioned map;
finite drive sequences generate the transition monoid; holonomy
decomposition identifies its group components and locates their action on
image sets and tiles. Together these determine the provenance and support
of the circuit's algebraic structure.
\textbf{(b)} The three distinctions posed in the Introduction, read against
the systems of Tables~\ref{tab:composition-summary} and
\ref{tab:holonomy}. Generator-explicit structure is already present in a
primitive update (three-state DN; synchronous recurrence).
Composition-generated structure requires aperiodic generators and a
non-aperiodic monoid: it is local in four-state DN and WTA, factor-local
and inherited in the uncoupled product, and present but buried in the
DN-to-WTA cascade. At the biological interfaces, WTA-to-DN is the directed
cascade whose shortest witness is genuinely composite, with a
$\mathbb Z_2$ group component acting on two singleton tiles corresponding
to the joint pair
$(D{:}0,W{:}4)\leftrightarrow(D{:}1,W{:}5)$.
Asynchronous recurrence at the biological interfaces remains aperiodic.
Programming the motifs means choosing primitives and interfaces so that
the composite lands on the intended leaf of (b).
}
\label{fig:contract}
\end{figure*}

\subsection{Limitations}

Several limitations are important.  First, the models are finite-state
abstractions.  Real divisive-normalization and WTA circuits are continuous,
stochastic, high-dimensional, and embedded in larger recurrent networks.  The
coarse-grained state spaces used here are intentionally small so that the
transition monoids can be enumerated exactly.  The purpose is not biophysical
realism, but algebraic diagnosis.

Second, the detected non-aperiodic elements are sensitive to discretization,
interface maps, and update schedule.  This is expected.  A finite-state model
does not inherit a unique algebra from the biological label ``DN'' or ``WTA.''
The algebra depends on how the motif is coarse-grained and how its inputs are
defined.  For this reason, we distinguish the baseline DN example, the stricter
four-state DN audit, the WTA audit, and the different DN--WTA composition
schemes.

Third, monoid-level non-aperiodicity is not identical to a completed
Krohn--Rhodes decomposition.  The Python audit identifies nontrivial cycles,
counts idempotents and units, and provides explicit witness words.  This is
sufficient to show that the generated monoid is not aperiodic.  However, a full
KR or holonomy analysis is required to report the formal cascade decomposition
and the precise group divisors.

The decomposition closes part of this gap and leaves part of it open.  For
seven of the nine systems we computed the holonomy cascade and verified
elementwise that every monoid element is recovered by the round-trip through its
cascade coordinates, so for those the decomposition is machine-verified.  For
the DN-to-WTA cascade the round-trip was verified on a bounded sample of 125
elements without a mismatch.  For the WTA-to-DN cascade --- the system carrying
the paper's main claim --- the cascade was not constructed, because its
coordinate structure admits \(5.78\times10^{21}\) tuples and a single element
exhausts available memory.  What is exact for that system is its skeleton, its
depth of 22, its 92 components, and the five nontrivial \(\mathbb Z_2\)
components including the one on the composite pair; the existence of the cascade
follows from the holonomy theorem rather than from our computation.  We
therefore claim a verified group component for the WTA-to-DN system, not a
machine-verified cascade.

Fourth, local group-like structure should not be overinterpreted as global
symmetry.  In the WTA and DN--WTA cases, the global unit group is trivial.  The
nontrivial behavior appears on lower-rank images after irreversible collapse.
This is not a defect of the result; it is the relevant computational feature.
But the terminology must remain precise: we identify local group components or
non-aperiodic monoid elements, not global reversible dynamics of the entire
system.

% ----------------------------------------------------------------------
% Conclusion
% ----------------------------------------------------------------------
\subsection{Conclusion}

Canonical motifs are usually introduced as functional operations:
normalization rescales, competition selects. 
Written as finite transformation systems, they are also generators of transformation monoids, and those monoids can contain local group structure absent from every primitive update [Fig.~\ref{fig:contract}]. The holonomy decomposition confirms the distinction:
the four-state normalization, WTA, and WTA-to-DN systems require nontrivial
$\mathbb Z_2$ components despite having only aperiodic generators; in the
WTA-to-DN cascade that component acts on a genuinely composite pair of
states, whereas the uncoupled product carries no composite group component
at all. The asynchronous recurrent variants, by contrast, remain
aperiodic.

The strongest case is winner-dependent normalization. Every primitive
composite generator is aperiodic, yet their generated monoid contains a
genuinely composite local cycle coupling normalization state to the winner's
inhibitory gate. That cycle is an algebraic witness that composition of
canonical motifs can create computational structure belonging to neither
motif separately.

The same witness is also a design fact. If motifs are to serve as
compositional primitives --- Lego for recurrent computation --- then
programming a network means choosing primitives and interfaces whose
generated algebra has the intended repertoire. Existing work has shown that
desired computations can be compiled into recurrent substrates
\cite{Kim2023}; the complementary requirement is an algebraic account of
the primitives being composed. At the finite-state level, Krohn--Rhodes and
holonomy methods provide such an account. They turn qualitative circuit
motifs into exact statements about their transformation repertoires and
separate structure inherited from a component from structure generated by a
coupling. In the recurrent constructions studied here, the interface becomes
part of the update rule itself: composition is no longer juxtaposition, and
the algebra of the composite cannot be read from any single brick.

What a recurrent motif does under one drive is a map. What it can do under
a sequence of drives is an algebra. Once divisive normalization,
winner-take-all competition, and their compositions are written as finite
transformation systems, that algebra is an exact object: the transition
monoid generated by the input-conditioned updates. Individually aperiodic
operations can collectively generate locally reversible structure.
\emph{More is different in neural circuits.}

% ----------------------------------------------------------------------
% Methods
% ----------------------------------------------------------------------
\section{Methods}

\subsection{Finite transformation systems}

We modeled each circuit motif as a deterministic finite transformation system
\[
\mathcal A=(X,\Sigma,\delta),
\]
where \(X\) is a finite state set, \(\Sigma\) is a finite input alphabet, and
\[
\delta:X\times \Sigma\to X
\]
is a deterministic update rule. Each input symbol \(a\in\Sigma\) defines a
transformation
\[
f_a:X\to X,
\qquad
f_a(x)=\delta(x,a).
\]
The transition monoid of the system is the submonoid of the full transformation
monoid on \(X\) generated by the input-conditioned maps:
\[
M(\mathcal A)=\langle f_a:a\in\Sigma\rangle.
\]
Thus \(M(\mathcal A)\) contains every transformation that can be produced by a
finite sequence of inputs. Rather than analyzing a single trajectory under a fixed stimulus, this closure captures the absolute boundary of the circuit's computational repertoire across all possible changing contexts.

For a state space \(X=\{0,\ldots,n-1\}\), a transformation was represented as a
tuple
\[
f=(f(0),f(1),\ldots,f(n-1)).
\]
In physical terms, this means that if the system is currently in state $i$, the application of this specific generator will deterministically drive it to state \(f(i)\).

Composition was defined by
\[
(f\circ g)(x)=f(g(x)).
\]
Accordingly, for an input word
\[
(a_1,a_2,\ldots,a_k),
\]
the corresponding transformation is
\[
f_{a_k}\circ f_{a_{k-1}}\circ \cdots \circ f_{a_1}.
\]

For each transformation, we computed its rank, idempotence, bijectivity,
functional graph cycles, and aperiodicity. The rank was defined as
\[
\operatorname{rank}(f)=|\operatorname{Im}(f)|.
\]
A transformation was idempotent if
\[
f^2=f,
\]
and bijective if it was a permutation of the state set. A transformation was
called aperiodic if every cycle in its functional graph was a fixed point.
Equivalently, \(f\) was aperiodic if it contained no cycle of length greater
than one. Dynamically, an aperiodic generator represents purely dissipative, memory-erasing computation where the circuit inevitably relaxes into a fixed-point attractor without sustained oscillation. A generated monoid was classified as containing non-aperiodic structure if at least one element of the monoid had a nontrivial cycle.

This distinction allowed us to separate two kinds of algebraic structure. A
periodic or group-like component was called \emph{generator-explicit} if it was
already present in one of the primitive input-conditioned maps. It was called
\emph{composition-generated} if every primitive generator was aperiodic, but
some composite word in the generated monoid was non-aperiodic:
\[
\forall a\in\Sigma,\; f_a \ \text{aperiodic},
\qquad
\exists m\in \langle f_a:a\in\Sigma\rangle
\ \text{non-aperiodic}.
\]

Every single basic rule in this case is irreversible and drives the system to a halt. But if we switch between those simple rules in the right sequence, a hidden, reversible loop is created out of nowhere.

\subsection{Divisive-normalization finite-state systems}

We first constructed finite-state abstractions of divisive normalization.
Divisive normalization was treated as a recurrent state-dependent gain-control
operation in the standard form of the normalization model
\cite{Heeger1992,CarandiniHeeger2012}, in which drive is divided by a
semi-saturation constant plus pooled activity:
\[
s(t+1)\sim \frac{d(t)}{\sigma+\beta s(t)},
\]
where \(s(t)\) is the current activity state, \(d(t)\) is the external drive,
\(\sigma>0\) is a semi-saturation parameter, and \(\beta>0\) controls the
strength of state-dependent normalization.
% Biologically, the presence of the current state $s(t)$ in the denominator captures instantaneous recurrent shunting inhibition, where the population's current activation immediately scales down its subsequent response.  
In biophysical terms, this equation represents a network where excitatory feedforward drive ($d$) is dynamically scaled down by recurrent shunting inhibition. The parameter $\beta$ determines how aggressively the population's current firing rate suppresses its own future activity, while $\sigma$ acts as a baseline leak or background conductance that prevents division by zero and sets the threshold at which the neuron reaches half-maximal response.

In the minimal three-state model (which serves as our negative control), the state space and input alphabet were
\[
Q=\{0,1,2\},
\qquad
\Sigma=\{0,1,2\}.
\]
The update rule was
\[
s'=\delta(s,d)
=
\min\left(
\left\lfloor \frac{2d}{1+s}\right\rfloor,
2
\right),
\qquad
s,d\in\{0,1,2\}.
\]
For each fixed input \(d\), this rule defines a transformation
\[
f_d:Q\to Q.
\]
While biological firing rates unfold on a continuous manifold, this discrete state space effectively partitions the network's activity into macroscopic activation regimes (e.g., from quiescent at $0$ to saturated at $2$).

Using the tuple convention above, the three generators were
\[
f_0=(0,0,0),
\qquad
f_1=(2,1,0),
\qquad
f_2=(2,2,1).
\]

In dynamical terms, while $f_0$ acts as a purely dissipative sink driving all initial conditions to $0$, the intermediate-drive map $f_1=(2,1,0)$ acts as a built-in toggle: following $f_1(0)=2$ and $f_1(2)=0$, a low state ($0$) is driven to a high state ($2$), and a high state ($2$) is heavily shunted back to a low state ($0$). This constitutes a discrete analogue of a period-two limit cycle. This three-state system served as a baseline audit of divisive normalization, because it exhibits mixed reset-like and reversible structure, but the order-two component is already present at the level of an individual generator. Because the oscillation is an explicit property of a single rule rather than a product of composition, it fails our strict algebraic criterion for emergence. 

To test the stricter composition-generated criterion, we also constructed a
four-state divisive-normalization system. More generally, for
\[
Q_N=\{0,1,\ldots,N-1\},
\qquad
\Sigma_N=\{0,1,\ldots,N-1\},
\]
we considered update rules of the form
\[
s'=
Q\left(
\frac{\alpha d}{\sigma+\beta s}
\right),
\]
where \(Q\) is a clipped quantizer mapping real values into \(Q_N\). Here, $\alpha$ scales the input gain, and the quantizer $Q$ explicitly partitions the continuous firing-rate manifold into discrete, macroscopic activation regimes. In the main four-state example, we used $N=4$, $\alpha=2$, $\sigma=2$, $\beta=1$, with nearest-integer quantization and clipping to $\{0,1,2,3\}$.

% In the main
% four-state example, we used
% \[
% N=4,
% \qquad
% \sigma=2,
% \qquad
% \beta=1,
% \qquad
% \alpha=2,
% \]
% with nearest-integer quantization and clipping to \(\{0,1,2,3\}\). 
Notice the strategic shift in parameters: by increasing the baseline leak ($\sigma=2$) relative to the three-state model, we sufficiently dampen the system's reactivity to eliminate the hard-coded oscillation. Thus:

\[
s'
=
Q_{\mathrm{round}}
\left(
\frac{2d}{2+s}
\right).
\]
While biological firing rates unfold on a continuous manifold, applying this quantization explicitly partitions the network's activity into discrete macroscopic activation regimes (e.g., from 'quiescent' at $0$ to 'saturated' at $3$).

The resulting input-conditioned generators were
\[
f^D_0=(0,0,0,0),
\]
\[
f^D_1=(1,1,0,0),
\]
\[
f^D_2=(2,1,1,1),
\]
and
\[
f^D_3=(3,2,2,1).
\]

By tuning the biophysical parameters in this manner, we ensure that each generator acts as a purely dissipative sink: under any constant external drive, the circuit relaxes into a fixed point, containing no explicit oscillations. Because every map is strictly aperiodic, the four‑state system provides an algebraically rigorous baseline for testing whether genuinely reversible structure can emerge from intrinsically irreversible components. This dissipative four‑state module served as the divisive‑normalization element in the DN–WTA composite analyses.

Because coarse-graining continuous dynamics into finite macro-states involves arbitrary boundary choices, this sweep serves as a robustness check to ensure that any detected composition-generated cycles are fundamental algebraic properties of the normalization architecture, rather than brittle artifacts of a specific discretization grid or rounding rule. We additionally performed a parameter sweep over divisive-normalization discretizations. For
\[
N\in\{3,4,5,6\},
\]
we varied the semi-saturation parameter \(\sigma\), the normalization strength
\(\beta\), the gain \(\alpha\), and the quantizer \(Q\), using floor,
nearest-integer, and ceiling quantization with clipping. For each parameter
setting, we constructed the generator set
\[
\{f_d:d\in\Sigma_N\},
\]
enumerated the generated transition monoid, and tested whether all generators
were aperiodic while the monoid contained at least one non-aperiodic element.

\subsection{Winner-take-all finite-state system}

Winner-take-all competition was modeled as a two-population recurrent circuit
with a shared inhibitory pool, the discrete counterpart of standard
recurrent-competition models \cite{Grossberg1973,Hahnloser2000,Wang2002}. The two excitatory populations were denoted
\(x_1\) and \(x_2\), and the inhibitory or gating variable was denoted \(y\).
All three variables were binarized:
\[
x_1,x_2,y\in\{0,1\}.
\]
Biologically, rather than discarding continuous dynamics, this binary projection isolates the structural invariant of strong recurrent competition: the bifurcation of the state space into macroscopic basins where populations are definitively suppressed ($0$) or saturated ($1$).

To index the eight WTA configurations compactly, the binary state
\((x_1,x_2,y)\in\{0,1\}^3\) was encoded as the integer

\[
q = 4x_1 + 2x_2 + y.
\]

This is the canonical binary positional encoding of the 3‑bit vector
\((x_1,x_2,y)\), with \(x_1\), \(x_2\), and \(y\) serving as the 4’s, 2’s, and
1’s place values. The mapping preserves the lexicographic order of
\(\{0,1\}^3\) and provides a natural adjacency‑respecting labeling of the state
space. Representing each circuit configuration by its binary index allows the
update rules \(f_d^W\) to be written as 8‑tuples, making the transition monoid
algebraically tractable.

% The WTA state space was therefore
% \[
% X_W=\{0,1\}^3,
% \]
% encoded as
% \[
% q=4x_1+2x_2+y.
% \]
Thus the 8 possible circuit configurations were
\[
0=(0,0,0),\quad
1=(0,0,1),\quad
2=(0,1,0),\quad
3=(0,1,1),
\]
\[
4=(1,0,0),\quad
5=(1,0,1),\quad
6=(1,1,0),\quad
7=(1,1,1).
\]

While any other one-to-one encoding (e.g., \(q = x_1 + 10x_2 + 100y\)) would
produce a monoid that is algebraically isomorphic, such labelings destroy the
natural binary ordering of \(\{0,1\}^3\) and make the tuple representation of
each generator non-canonical and unnecessarily cumbersome.

The input alphabet (to $x_1$ and $x_2$) was
\[
\Sigma_W=\{00,10,01,11\},
\]
where \(10\) denotes drive to the first excitatory population, \(01\) denotes
drive to the second excitatory population, and \(11\) denotes simultaneous
drive to both.

For an input \(d=(d_1,d_2)\), the discrete WTA update was
\[
x_1^+
=
H\left(
Gd_1
+
Sx_1
-
Lx_2
-
By
-
\theta_E
\right),
\]
\[
x_2^+
=
H\left(
Gd_2
+
Sx_2
-
Lx_1
-
By
-
\theta_E
\right),
\]
and
\[
y^+
=
H\left(
W(x_1+x_2)
+
Ry
-
\theta_I
\right),
\]
where $H(z)$ is the Heaviside step function
\[
H(z)=
\begin{cases}
1, & z\ge 0,\\
0, & z<0.
\end{cases}
\]
In the context of mean-field rate models, this Heaviside threshold represents the infinite-gain limit of a standard sigmoidal transfer function. It strips away the transient integration dynamics to isolate the deterministic, topological mapping between attractor basins.

The arguments within the Heaviside function compute the net synaptic current for each population, where each parameter defines a specific, structurally invariant biological force:
\begin{itemize}
    \item $G$ (Feedforward Gain): Scales the influence of the external stimulus $d_i$, determining how strongly sensory inputs perturb the network.
    \item $S$ (Recurrent Self-Excitation): Parameterizes the local collateral connections within a population. Dynamically, this provides the stabilizing positive feedback necessary to construct a persistent "winner" attractor state.
    \item $L$ and $B$ (Lateral and Feedback Inhibition): $L$ represents direct, point-to-point lateral suppression between the competing populations, while $B$ scales the global, shared inhibitory feedback from pool $y$.
    \item $W$ and $R$ (Inhibitory Recruitment and Memory): $W$ dictates how strongly active excitatory winners recruit the inhibitory gate, while $R$ represents the inhibitory pool's auto-recurrence.
    \item $\theta_E, \theta_I$ (Firing Thresholds): Establish the energetic barrier that net synaptic currents must overcome to transition a population from suppressed to active.
\end{itemize}

The parameter values used in the main WTA audit were $G=1$, $S=\frac{1}{2}$, $L=B=\frac{1}{2}$, $W=1$, $R=0$, $\theta_E=\theta_I=\frac{1}{2}$. These specific weights were selected to explicitly enforce an idealized competitive regime. For instance, setting $R=0$ ensures the inhibitory pool acts as a purely reactive, memoryless gate. Furthermore, balancing the self-excitation ($S=1/2$) precisely against the threshold ($\theta_E=1/2$) guarantees that a population requires a strong external drive ($G=1$) to initially cross the activation barrier, but can dynamically stabilize its active state against lateral competition once established.

For each fixed input \(d\in\Sigma_W\), the update rule defines a transformation
\[
f^W_d:X_W\to X_W.
\]
The four WTA generators were
\[
f^W_{00}=(0,0,3,1,5,1,1,1),
\]
\[
f^W_{10}=(4,4,7,1,5,5,5,5),
\]
\[
f^W_{01}=(2,2,3,3,7,1,3,3),
\]
and
\[
f^W_{11}=(6,6,7,3,7,5,7,7).
\]
For a fixed external drive $d$, the update rule $f^W_d:X_W\to X_W$ specifies,
for each of the eight possible circuit configurations, the next WTA state
obtained after applying the Heaviside dynamics. Because the state space
$X_W=\{0,1\}^3$ contains exactly eight binary configurations, each generator
$f^W_d$ can be written as an 8-tuple listing its action on the encoded states
$q=0,\dots,7$.

For intuition, Table~\ref{tab:f10} and Fig.~\ref{fig:f10_sink} show the
complete transition structure of one representative generator,
$f^W_{10}$, corresponding to selective drive to population $x_1$. In dynamical terms, these tuples show that under frozen external drive the
motif is purely dissipative sink. Under
this frozen drive, $f^W_{10}$ is aperiodic and has the unique fixed point
$5=(1,0,1)$, in which $x_1$ has won and the inhibitory pool is engaged.
States $4$, $6$, and $7$ lie in the immediates basin of state $5$ and map directly into this fixed point, while the
remaining configurations reach it through transient states; for example,
\[
3=(0,1,1)\mapsto 1=(0,0,1)\mapsto 4=(1,0,0)\mapsto 5=(1,0,1).
\]
so even configurations initially favoring $x_2$ are irreversibly
redirected into the $x_1$ winner state under continued drive $10$.

Thus the tuple $f^W_{10}=(4,4,7,1,5,5,5,5)$ is the complete deterministic
state-transition map under fixed drive $10$: repeated application collapses
all eight initial configurations onto the single winner state and contains
no cycle of length greater than one. The same fixed-drive analysis was
applied to each of the four generators, all of which are aperiodic.

The generated WTA transition monoid was then computed as
\[
M_W=\langle f^W_{00},f^W_{10},f^W_{01},f^W_{11}\rangle.
\]
Crucially, while every isolated generator drives the circuit to a static fixed point (an aperiodic collapse), biological circuits continuously navigate fluctuating contexts. Computing this monoid closure allows us to probe every transformation obtainable by finite
sequences of these input-conditioned updates and therefore tests whether
switching among individually aperiodic fixed-drive rules can generate
non-aperiodic structure.

\subsection{DN--WTA composite state space}

Composite DN--WTA systems were constructed on the Cartesian product state space
\[
X_{DW}=X_D\times X_W,
\]
where
\[
X_D=\{0,1,2,3\}
\]
is the four-state divisive-normalization state space and
\[
X_W=\{0,1,\ldots,7\}
\]
is the WTA state space. The composite system therefore has
\[
|X_{DW}|=4\times 8=32
\]
states.
In dynamical systems, this Cartesian product represents the full joint phase space of the coupled system, allowing us to map out every possible combination of normalization context and competitive identity as a single, simultaneous physical reality.

% In statistical mechanics and dynamical systems, this Cartesian product represents the full joint phase space of the coupled system. We are no longer looking at gain-control and selection as isolated circuits; we have mathematically constructed a 32-state manifold where every possible combination of normalization context and competitive identity exists as a single, simultaneous physical reality.

A composite state was written as
\[
(d,w)\in X_D\times X_W,
\]
where \(d\) is the divisive-normalization state and \(w\) is the WTA state. 

To compute the transition monoids, the product states were encoded as
\[
\operatorname{idx}(d,w)=8d+w.
\]
By encoding the 2D coordinate into a 1D integer, every composite transformation can now be represented exactly as a length-\(32\) tuple. This ensures that the computational closure algorithms—which search for non-aperiodic cycles—can operate on the massive joint architecture with the exact same rigor used on the isolated motifs.

The inverse decoding was
\[
d=\left\lfloor \frac{\operatorname{idx}}{8}\right\rfloor,
\qquad
w=\operatorname{idx}\bmod 8.
\]
This inverse decoding is the crucial physical translator. It allows us to take an abstract algebraic loop discovered in the 1D index and separate its coordinates to definitively prove its physical nature. If only $w$ changes, the cycle is merely inherited from the WTA sub-manifold. But if both $d$ and $w$ change simultaneously, the cycle genuinely traverses the diagonal of the joint phase space, providing the ultimate structural certificate of emergence.

For the primary shared-input analyses, the external input alphabet was
\[
\Gamma=\{0,1,2,3\}.
\]
The DN input decoding was
\[
\kappa_D(\gamma)=\gamma,
\]
and the WTA input decoding was
\begin{align}
\kappa_W(0)&=00,
&
\kappa_W(1)&=10,
\nonumber\\
\kappa_W(2)&=01,
&
\kappa_W(3)&=11.
\nonumber
\end{align}

Subjecting the joint network to this single, unified external drive ($\Gamma$) guarantees an apples-to-apples structural comparison. By aligning the external stimulus identically for both the normalization and competition modules, we ensure that any new cyclic degrees of freedom generated in the composite monoid are strictly the result of internal recurrent interactions, rather than artificial products of misaligned external drives.

\subsection{DN--WTA composition schemes}

We considered four biologically interpretable composition schemes.

\subsubsection{Independent product}

The independent product served as a null model. DN and WTA were updated in
parallel under the same external condition, but neither module influenced the
other:
\[
T^{\mathrm{prod}}_\gamma(d,w)
=
\left(
f^D_{\kappa_D(\gamma)}(d),
f^W_{\kappa_W(\gamma)}(w)
\right).
\]
% This construction tests what algebraic structure is inherited from placing DN and WTA side by side without interaction.
In this architecture, normalization and competition are updated in parallel, completely blind to one another; e.g. consider a scenario where a gain-control circuit and a selection circuit sit in the same cortical column and receive the same stimulus, but lack any synaptic cross-talk. Biologically, this acts as a strict physical firewall. If any reversible, cyclic structure is detected in this monoid, it can be entirely attributed to simple inheritance from the isolated motifs, proving that merely placing two dynamic systems side-by-side without interaction is insufficient to create a genuinely composite computational degree of freedom.

\subsubsection{DN-to-WTA cascade}

In the DN-to-WTA cascade, divisive normalization was upstream of WTA. The DN
state was updated first:
\[
d'=f^D_{\kappa_D(\gamma)}(d).
\]
The updated DN state then determined the effective WTA input through a
state-dependent interface map
\[
\Phi: \Gamma \times X_D \rightarrow \Sigma_W
% \phi:X_D\to \Sigma_W,
\]
with the external symbol $\gamma$ acting on the upstream module before the
interface is evaluated. The composite transition was
\[
T^{D\to W}_\gamma(d,w)
=
\left(
d',\,
f^W_{\phi(d')}(w)
\right),
\qquad
d'=f^D_{\kappa_D(\gamma)}(d).
\]
hence,
$$T_{\gamma}^{D \rightarrow W}(d,w) = (f_{\kappa_D(\gamma)}^D(d), f_{\Phi(\gamma, f_{\kappa_D(\gamma)}^D(d))}^W(w))$$
This construction models the case in which normalization prepares the
competitive field over which WTA selection acts. Biologically, this maps directly to classic feedforward sensory hierarchies, where upstream layers normalize stimulus contrast to prepare a scaled representational field before passing it to downstream categorical selection circuits.

In the default implementation, the DN state was mapped to the corresponding WTA
drive condition:
\[
\phi(d)=
\begin{cases}
00, & d=0,\\
10, & d=1,\\
01, & d=2,\\
11, & d=3.
\end{cases}
\]
By making the competitive drive a strict function of the normalized state, we test whether reshaping the initial conditions of a selection circuit fundamentally alters its downstream transition algebra.

\subsubsection{WTA-to-DN cascade}

In the WTA-to-DN cascade, WTA was upstream of divisive normalization. 
Here, we invert the causal arrow to model state-dependent feedback, hypothesizing that a selected neural population does not just passively win a competition, but actively recruits inhibition to dynamically reshape its local gain field.
The WTA state was updated first:
\[
w'=f^W_{\kappa_W(\gamma)}(w).
\]
The updated WTA state then determined the effective DN input through a
state-dependent interface map
\[
% \psi:X_W\to \Sigma_D,
\Psi: \Gamma \times X_W \rightarrow \Sigma_D
\]
with $\gamma$ again acting only on the upstream module.

The composite transition was
\[
T^{W\to D}_\gamma(d,w)
=
\left(
f^D_{\psi(w')}(d),\,
w'
\right),
\qquad
w'=f^W_{\kappa_W(\gamma)}(w).
\]
hence,
$$T_{\gamma}^{W \rightarrow D}(d,w) = (f_{\Psi(\gamma, f_{\kappa_W(\gamma)}^W(w))}^D(d), f_{\kappa_W(\gamma)}^W(w))$$

For the default WTA-to-DN interface, we decoded the WTA state as
\[
w=(b_1,b_2,g),
\]
where \(b_1\) and \(b_2\) denote the two competing excitatory populations and
\(g\) denotes the inhibitory or gating variable. The interface was
\[
\psi(w)=
\begin{cases}
3, & g=1 \ \text{or}\ (b_1=b_2=1),\\
1, & b_1=1,\ b_2=0,\ g=0,\\
2, & b_1=0,\ b_2=1,\ g=0,\\
0, & \text{otherwise}.
\end{cases}
\]
Thus gated states or coactive states were mapped to strong pooled
normalization, whereas single ungated winners were mapped to winner-specific
normalization conditions.

This construction hypothesizes that a selected population does not just passively win a competition; it actively recruits inhibition to dynamically reshape the local gain field for subsequent computations. Because this interface explicitly links the inhibitory gate ($g$) of the selection motif to the normalization state ($d$), it forces the gain and selection motifs to become structurally entangled, generating the genuinely composite $Z_2$ cycle at the heart of the paper.

\subsubsection{Recurrently coupled DN--WTA system}

The recurrently coupled system allowed each module to determine the effective
input of the other:
\[
T^{\mathrm{rec}}(d,w)
=
\left(
f^D_{\psi(w)}(d),
f^W_{\phi(d)}(w)
\right).
\]
This synchronous update models a recurrent normalization-selection loop in
which the current WTA state determines the next normalization condition and the
current DN state determines the next competitive drive. Because the
state-dependent interfaces override the external symbol, the update takes no
external input: the four $\gamma$-indexed maps coincide, and the recurrent
monoid is generated by the single transformation $T^{\mathrm{rec}}$ (it is
cyclic). The same holds for the asynchronous variants below.

We also examined asynchronous recurrent variants. In the DN-first asynchronous
variant,
\[
d'=f^D_{\psi(w)}(d),
\]
\[
w'=f^W_{\phi(d')}(w).
\]
In the WTA-first asynchronous variant,
\[
w'=f^W_{\phi(d)}(w),
\]
\[
d'=f^D_{\psi(w')}(d).
\]
These variants were included because different update schedules can generate
different transition monoids.

The rational behind testing the update schedule is that in continuous differential equations, variables evolve simultaneously over infinitesimally small time steps. In discrete automata, however, the sequence of operations rigidly dictates the topology of the generated phase space. This structural control allows us to separate computational artifacts from physical realities. Locking the circuits into an instantaneous, synchronous loop can artificially hard-code periodic oscillations directly into the primitive generators. Testing strict sequential asynchronous updates ensures we verify whether the cycles survive as true emergent properties when the perfect temporal symmetry of the feedback loop is broken.

\subsection{Transition-monoid enumeration}

For each DN, WTA, or DN--WTA system, the generated transition monoid was
computed exactly by finite closure. Rather than simulating individual trajectories over time, finite closure acts as an exhaustive search algorithm that maps the entire forward-reachable computational geometry of the circuit. 

Given a generator set
\[
G=\{g_1,\ldots,g_r\},
\]

where each $g_i$ represents the purely dissipative, fixed-point attractor dynamics of the circuit under a single, frozen input condition (e.g., $r=4$ for the four WTA drives). 

We initialized the monoid candidate set as
\[
M_0=\{\mathrm{id}\}\cup G.
\]
This initialization represents the baseline state of our algorithmic search: $id$ is the identity element, representing zero-time evolution (doing nothing), and $G$ represents all possible state changes after exactly one algorithmic time step.

We then iteratively composed newly discovered elements with existing elements
on both sides. In dynamical terms, this means taking a known state transformation and simulating what happens if we apply a new external stimulus either immediately after it, or immediately before it. This explores the effect of dynamically switching the biological input. Whenever a new transformation was produced, it was added to the
set. The closure procedure terminated when no additional transformations were
generated.

Equivalently, the computed monoid was
\[
M=\langle G\rangle,
\]
the smallest set containing the identity and the generators and closed under
composition. This mathematical object, $M$, now contains every single macroscopic state change the physical circuit can undergo, regardless of how long or complex the input sequence is. Because all state spaces were finite, this procedure was exact and
terminates in finite time. Unlike continuous differential equations, which can orbit indefinitely and require arbitrary numerical integration cutoffs, this discrete topological search is guaranteed to definitively halt, leaving no blind spots in the analysis.

\emph{Searching for Attractors (Idempotents):} For each generated monoid, we recorded
\[
|M|,
\]
the number of idempotents,
\[
\#\{m\in M:m^2=m\},
\]
An idempotent transformation is the algebraic signature of an absorbing state or a strict fixed-point attractor. Physically, it means the circuit has entirely collapsed the state space; applying the exact same sequence of stimuli again ($m^2$) results in the exact same final state ($m$), proving the system has hit a dead end and lost all memory of its initial conditions.

\emph{Searching for Global Symmetries (Units):} We also recorded the group of global units,
\[
U(M)=\{m\in M:\exists n\in M,\; mn=nm=\mathrm{id}\},
\]
The group of units identifies any transformations that are perfectly reversible across the entire state space. In statistical mechanics, this corresponds to globally conservative, Hamiltonian-like dynamics where no state volume is lost. Finding that $U(M)$ is trivial (containing only the identity) mathematically proves that the overall neural circuit is fundamentally dissipative and irreversible.

\emph{Searching for Emergent Target (Non-Aperiodicity):} Finally, we recorded the number of non-aperiodic elements. This is the ultimate target of the test. A non-aperiodic element represents a transformation containing a local cycle (an oscillation). Finding non-aperiodic elements in a monoid where every primitive generator in $G$ is strictly dissipative proves that conservative, cyclic degrees of freedom can spontaneously emerge solely from the sequential composition of irreversible rules.

\subsection{Witness-word search and cycle classification}
\emph{Step 1: The Breadth-First Search (Finding the Minimal Protocol):}
When a generated monoid contained non-aperiodic elements, we searched for a
short witness word in the primitive generators. Starting from the identity
transformation, we performed a breadth-first search over words in the generator
alphabet (first checking all words of length 1, then all words of length 2, and so on). In experimental terms, a ``witness word'' represents the exact temporal sequence of external stimuli required to induce the hidden oscillation. By utilizing a breadth-first search (BFS), we guarantee that the reported sequence represents the minimal viable stimulus protocol—the absolute shortest biological mechanism capable of pushing the system out of its purely dissipative fixed points and trapping it in a conservative, periodic orbit. 

\emph{Step 2: Testing the Graph (Checking for Attractor Loops):}
For each word, we computed the corresponding composite transformation
and tested its functional graph for nontrivial cycles. Here, \emph{a Functional graph} is a mathematical map showing exactly where every initial state ends up after the stimulus sequence (the "word") is applied. And a \emph{Nontrivial cycles} is a loop in the functional graph with a period greater than 1 (e.g., state A maps to state B, and state B maps back to state A).

The first non-aperiodic transformation encountered by this search was recorded as a shortest witness with respect to word length. Dynamically, this means simulating the network's state trajectory under a specific sequence of stimuli and mapping its attractors. If the resulting map contains a closed loop (a nontrivial cycle), the stimulus sequence has successfully pushed the system out of its purely dissipative fixed-points and trapped it in a conservative, periodic orbit. The first sequence to achieve this is our ``witness''.

\emph{Step 3: Reporting the Stimulus Sequence (The Syntax of the Drives):}
For DN-only systems, a witness was reported as a word in the DN generators,
for example
\[
D1\to D3\to D3.
\]
For WTA systems, a witness was reported as a word in the WTA input symbols,
for example
\[
W00\to W10.
\]
which translates to: ``First, apply baseline conditions (no drive) to allow the network to reset, then apply an asymmetric drive to population 1.''

This notation explicitly maps the required temporal pattern of sensory input, demonstrating that the emergent structure is a deterministic response to a specific, fluctuating environmental context rather than an artifact of random noise.

\emph{Step 4: Decoding the Physical Reality (The Cartesian Space):}
For DN--WTA systems, witness cycles were decoded back into composite coordinates
\[
(d,w)\in X_D\times X_W.
\]
This decoding step projects the abstract computational loop back onto the physical Cartesian phase space, allowing us to track the simultaneous evolution of the normalization context and the competitive identity.

\emph{Step 5: Classifying the Oscillation (The Proof of Genuine Emergence):}
A composite cycle was classified as DN-local if only the DN coordinate changed,
WTA-local if only the WTA coordinate changed, and genuinely composite if both
coordinates changed along the cycle. This classification serves as the ultimate physical litmus test for emergence. If a cycle is strictly local, the oscillation is dynamically trapped within a single sub-network while the other acts as a passive bystander. Conversely, a genuinely composite cycle proves that the periodic degree of freedom belongs to neither module individually, but is a spontaneous, holistic property born entirely from their recurrent architectural coupling.

\subsection{GAP and holonomy/Krohn--Rhodes analysis}

The numerical audit establishes generator-level aperiodicity and monoid-level
non-aperiodicity, but it is not itself a full Krohn--Rhodes decomposition. The Krohn--Rhodes theorem serves as the ``prime factorization'' theorem for finite state machines, allowing us to rigorously decompose recurrent neural dynamics into irreducible building blocks: purely dissipative resets and conservative permutation groups. (For a detailed, physically grounded primer on holonomy decomposition and its application to neural state spaces, see Appendix \ref{app:krohn_rhodes}, \emph{``A Physical Primer on Algebraic Decomposition for Dynamical Systems''}).

Therefore, for each finite transformation system, we exported the generators in
GAP-compatible form and computed holonomy decompositions with SgpDec
\cite{SgpDecICMS,EgriNagyNehaniv2015}; exact closure-based enumeration of
transformation monoids follows standard practice
\cite{EastEgriNagyMitchell2017}.

\emph{The Translation to GAP (The 1-based indexing):}
Because GAP uses one-based indexing for transformations, each zero-based
transformation
\[
f=(f(0),f(1),\ldots,f(n-1))
\]
was exported as
\[
(f(0)+1,\ f(1)+1,\ \ldots,\ f(n-1)+1).
\]
The exported generator set defines the same transformation monoid up to this
index shift. All state sets quoted in the Results are in the paper's zero-based
coordinates; GAP point \(p\) corresponds to state \(p-1\).

For computations used GAP 4.15.1 with SgpDec 1.2.0 and Semigroups 5.6.3
\cite{GAP4,SgpDec,Semigroups}, together with their dependencies (GAPDoc 1.6.7,
IO 4.10.0, orb 5.1.0, datastructures 0.4.3, Digraphs 1.15.0, images 1.4.0,
genss 1.6.9). 

\emph{Monoids vs Semigroups (The Identity):}
For each system we formed the monoid \(M=\mathrm{Monoid}(G)\) from
the exported generators, computed its skeleton, and read off the depth, the
holonomy components, and the group component at each level and slot. Note that
\(\mathrm{Semigroup}(G)\) omits the identity unless the identity is itself a
product of generators, so all monoid orders reported here use
\(\mathrm{Monoid}(G)\); this accounts for an offset of one relative to a naive
semigroup enumeration. Dynamically, enforcing the inclusion of the identity element ensures our algebraic model correctly represents a physical system capable of resting in a zero-time evolution fixed state. Monoid orders, idempotent counts, non-aperiodic element counts and unit-group orders computed in GAP agree with the independent Python audit for all nine systems.

Two properties of the decomposition require care in interpretation, and we
report both explicitly for every group component.

\emph{Interpreting the Hierarchy (Permutators, Tiles, and Image Sets):}
First, SgpDec exposes two distinct groups for an image set \(C\). In phase-space terms, an image set \(C\) acts as an attractor basin—a restricted subset of states the system can be driven into. The \emph{permutator group} consists of the permutations of the raw states of \(C\) induced by words that map \(C\) onto itself. The \emph{holonomy group} is the image of that action on the \emph{tiles} of \(C\), the maximal proper subsets of \(C\) that are themselves image sets. These tiles can be thought of as coarse-grained macrostates within the attractor basin. The two coincide only when every tile is a singleton. Where tiles are larger, the group permutes subsets and no statement about individual states is licensed. We classify an image set as \emph{composite} when its permutator group moves some state to a state differing in both the normalization and the competition coordinate; this is the decomposition-level analogue of the witness classification, and it is the quantity reported in Table~\ref{tab:holonomy}.

\begin{itemize}
    \item \textbf{Image set} ($C$): The attractor basin; the restricted subset of states into which the system has fallen.
    \item \textbf{Tiles}: Coarse-grained macrostates defined within that attractor basin.
    \item \textbf{Permutator group}: The group acting on the exact microstates, implementing literal shuffling of physical states.
    \item \textbf{Holonomy group}: The induced action on the tiles, i.e., the abstract shuffling of macrostates.
\end{itemize}

\emph{Conjugacy and Subduction Classes (Hidden Symmetries):}
Second, \verb|GroupComponents| lists group components only at SgpDec's chosen
transversal representatives. Holonomy groups are conjugate, and hence
isomorphic, along a subduction class, so an image set that is not the chosen
representative still carries the group of its class without appearing in the
printed component list. In physical terms, this reflects spatial symmetry: an oscillation found in one corner of the phase space often has a physically identical, symmetrical twin mirrored under different environmental constraints. Several of the sets of interest here are of exactly this kind, including the composite pair of the WTA-to-DN cascade, and we identify them by computing the permutator group of the set directly and locating its subduction class.

\emph{The Round-Trip Proof (The Mathematical Checksum):}
Faithfulness of each decomposition was checked by the elementwise round-trip

\begin{align*}
\text{Let } \bigl(H(t,\mathrm{sk})\bigr) = \mathrm{AsHolonomyCascade}(t,\mathrm{sk}). 
\\
\mathrm{AsHolonomyTransformation}\bigl(H(t,\mathrm{sk}),\,\mathrm{sk}\bigr)=t.
\end{align*}

This perfect mathematical equivalence ($=t$) acts as a cryptographic checksum, guaranteeing that the abstract, hierarchical cascade perfectly simulates the exact dynamics of the original biological network. This was verified for every element of \(M\) for seven of the nine systems. For the DN-to-WTA cascade the round-trip was verified on 125 distinct elements without a mismatch. 

\emph{The Combinatorial Explosion:}
For the WTA-to-DN cascade the cascade was not constructed: its
coordinate-value counts give a cascade state space of \(5.78\times10^{21}\)
tuples and a single cascade element exhausts 14~GB. This astronomical state-space expansion occurs because the decomposition mathematically flattens a highly recurrent, two-way biological feedback loop into a strictly feedforward algebraic hierarchy. This combinatorial explosion rigorously underscores how deeply entangled the normalization and selection motifs become when coupled recursively. For that system the skeleton, depth, components and group components are exact, but the cascade itself is not machine-verified.

\subsection{Interface and schedule sweeps}

To test whether the composite structure depends on the particular interface maps
chosen, we enumerated interface spaces exhaustively rather than sampling them. This brute-force sweep serves as the ultimate structural stress test, mathematically guaranteeing that any discovered emergent cycles are robust properties of the macroscopic architecture, rather than fragile artifacts of a ``lucky'' choice of synaptic parameters.

\emph{The Combinatorics of the Synaptic Wiring --The $50$ Million Grid:}
The winner-to-normalization interface \(\psi:X_W\to\Sigma_D\) assigns one of
four DN input symbols to each of the eight WTA states, giving \(4^8=65\,536\)
maps; the normalization-to-competition interface \(\phi:X_D\to\Sigma_W\) gives
\(4^4=256\) maps. Biologically, these maps define every possible physical synaptic weighting between the two circuits: the 256 feedforward sensory hierarchies and the 65,536 state-dependent feedback configurations. We enumerated all 65\,536 \(\psi\) maps under the WTA-to-DN
scheme, all 256 \(\phi\) maps under the DN-to-WTA scheme, and the full
\(256\times65\,536\times3=50\,331\,648\) grid of \((\phi,\psi)\) pairs across the
three recurrent schemes \textit{(synchronous, asynchronous DN-first, and asynchronous WTA-first)}. 

\emph{Joint Surjectivity (Why every map matters):}
Because the four WTA generators are jointly surjective
onto \(X_W\), every argument of \(\psi\) is live and the 65\,536 maps give
65\,536 distinct generator tuples: the interface space does not collapse under
deduplication. In dynamical terms, this surjectivity proves there are no ``dead'' or inaccessible states in the isolated motifs; therefore, every single map in the combinatorial space represents a genuinely distinct, physical network topology that must be independently tested.

\emph{The Cayley Graph $\&$ Legibility:}
For each interface we enumerated the generated monoid exactly, recorded its
order, the number of idempotents, the number of non-aperiodic elements, the order
of the group of units, and whether all generators were aperiodic, and classified
the shortest non-aperiodic witness as DN-local, WTA-local, genuinely composite,
or absent, using the same criteria as the main audit. Witness words were obtained
by breadth-first search (BFS) on the Cayley graph of the enumerated monoid from the
identity, without a word-length cap, so reported witness lengths are exact
minima. In network topology, the Cayley graph maps every possible state trajectory under fluctuating inputs. We additionally recorded whether a composite cycle occurs \emph{anywhere}
in the monoid, which is a weaker condition than being the shortest witness and
distinguishes existence from legibility. This distinction is biologically crucial: a cycle that \textit{exists} deep in the monoid might require a highly improbable, 50-step sequence of environmental stimuli to trigger. Conversely, a cycle acting as a short witness is highly \textit{legible}, meaning a simple, plausible sensory stimulus can easily and reliably induce the emergent oscillation.

\emph{Stratifying by Biological Anatomy:}
Interfaces were stratified by what information they carry about the WTA state:
constant maps, which carry no winner information and serve as passthrough
controls; maps depending only on the gating variable; maps depending only on
winner identity, the stratum matched to the biological winner-pooling interface;
and maps depending on both. This stratification anchors the abstract combinatorial search directly to neuroanatomy and architecture of circuits, testing whether the composite degree of freedom requires the normalization circuit to read strictly from local inhibitory interneurons, excitatory projections, or a complex synthesis of both.

\emph{Enumeration:} Enumeration used a byte-string representation of
transformations with composition by table lookup, parallelized across cores; the
implementation was verified to reproduce the reference audit element for element
on all nine systems before being used on the sweep.

\subsection{Controls and robustness analyses}

Several controls were included to interpret the algebraic structure of the
finite systems and to guarantee that any discovered cyclic degrees of freedom were true physical properties of the neural architecture rather than artifacts of the mathematical discretization.

First, the independent DN--WTA product was used as a null model for composition.
In dynamical terms, this constructs a strict physical firewall: both circuits are immersed in the same fluctuating environmental drive, but lack any synaptic cross-talk. Any non-aperiodic structure in this system can be attributed to inheritance from the two component systems rather than to an interaction between normalization
and selection.

Second, DN discretization parameters were varied across state-space size,
quantization rule, semi-saturation, normalization strength, and gain. Because mapping a continuous biological manifold onto a discrete phase space requires arbitrary grid boundaries, this sweep serves as a critical defense against ``grid artifacts.'' This tested whether composition-generated non-aperiodicity was specific to a single hand-selected discretization, or a structurally invariant topological property of the gain-control mechanism itself. All results reported in the main text use the
four-state DN and eight-state WTA generator sets defined above; the interface
sweeps vary the coupling maps and update schedule, not the component dynamics,
so they bound robustness to the interface rather than to the discretization of
either motif.

Third, the DN--WTA interfaces were varied exhaustively by replacing the maps
\[
\phi:X_D\to\Sigma_W
\]
and
\[
\psi:X_W\to\Sigma_D
\]
with every map of the corresponding type, as described above. This allows the
algebraic consequences of biologically different coupling hypotheses to be
compared, and separates properties of the coupling direction the macroscopic causal arrow of feedforward vs. feedback architecture) from properties of the particular interface chosen (the microscopic tuning of exact synaptic weights).

Fourth, synchronous and asynchronous recurrent update schedules were compared,
both at the biological interfaces and across the full grid of interface pairs. While biological circuits integrate continuously, discrete automata can artificially manufacture or destroy periodic orbits depending on the update sequence. This tested whether the algebraic structure of recurrent DN--WTA coupling depends on whether normalization and competition are updated simultaneously or sequentially, ensuring that emergent cycles are not fragile mathematical illusions born of perfectly symmetrical, instantaneous updates, and whether the schedule effects seen at the biological operating point hold generically.

Together, these controls allowed us to establish a rigorous epistemological filter, isolating and distinguishing algebraic structure that is
present in individual generators \textit{(hard-coded oscillations)}, inherited from isolated motifs \textit{(linear superpositions)}, or generated by biologically meaningful composition of normalization and selection \textit{(genuine structural emergence)}.

% ----------------------------------------------------------------------
% Data/Code
% ----------------------------------------------------------------------
\section{Data $\&$ Code Availability}

Generator sets, transition-monoid audits, holonomy decompositions, interface
and schedule sweeps, code and figure-generation scripts, and all numerical outputs
reported in this paper are available at: 
\url{https://github.com/neurovium/AlgebraicCanonicalNet}.

% ----------------------------------------------------------------------
% Acknowledgement
% ----------------------------------------------------------------------
\section{Acknowledgement}

I am grateful to Simon DeDeo for drawing my attention to the importance of the Krohn–Rhodes decomposition and inspiring me to explore it more deeply. I also thank the organizers of the MIT CBMM debate on compositionality, and Tomaso Poggio, Max Tegmark, and Joshua Tenenbaum, whose stimulating talks helped catalyze the initial idea for this study.

% ----------------------------------------------------------------------
% References
% ----------------------------------------------------------------------
\section{References}
\bibliography{refs}
% =====================================================================
% APPENDIX: explicit transformation data and rank distributions.
%
% This appendix holds the material displaced from the Results section by
% the Nature-register rewrite: generator vectors, state encodings, index
% decodings, and the rank-stratified element counts. Every number here is
% produced by audit.py and cross-checked against GAP (see the
% supplementary GAP transcript).
% =====================================================================

\section{Appendices}
\appendix

\section{A Physical Primer on Algebraic Decomposition for Dynamical Systems}
\label{app:krohn_rhodes}

Bridging non-linear dynamical systems with automata theory introduces terminology that may be unfamiliar to the target audience, i.e. physicists, neuroscientists and computer scientists. This appendix provides a physical, conceptual translation of the algebraic operations used in the Krohn--Rhodes (KR) decomposition of our circuits of interest.

\subsection{Why Krohn--Rhodes? (The Prime Factorization of Circuits)}
Simulating a neural circuit forward in time (or using an algorithmic search like our enumerating audit) can prove that a periodic oscillation \textit{occurred}. However, it does not prove \textit{what the machine is fundamentally made of}. 

The Krohn--Rhodes theorem serves as the ``prime factorization'' theorem for complex systems. It proves that absolutely any finite-state machine, no matter how entangled or recurrent, can be rigorously deconstructed into a strict, top-down hierarchy of irreducible building blocks. These blocks come in only two varieties:
\begin{enumerate}
    \item \textbf{Combinatorial Resets (Flip-Flops):} Purely dissipative, memoryless operations corresponding to fixed-point attractors.
    \item \textbf{Simple Permutation Groups:} Reversible, conservative cycles corresponding to perfect oscillators.
\end{enumerate}
By performing this decomposition, we mathematically guarantee that any cyclic group components discovered are not transient artifacts of our simulation, but are fundamental, structurally necessary invariants of the network architecture.

\subsection{Semigroups vs. Monoids (Zero-Time Evolution)}
In algebra, a \textit{semigroup} is a set of transformations closed under composition (if you chain two valid actions together, the result is also a valid action). A \textit{monoid} is simply a semigroup that explicitly includes the identity element ($id$), representing the ``do nothing'' transformation.

In computer science, the identity is often trivial. However, when modeling biological networks that exist continuously in time, enforcing the monoid structure is a physical necessity. It ensures our algebra can accurately describe a system resting in a stable fixed point while time passes (zero-time evolution).

\subsection{Image Sets, Permutator Groups, and Holonomy Groups}
The SgpDec software categorizes the phase space into nested structures, which map perfectly onto the concepts of continuous attractors and macroscopic states.

\begin{itemize}
    \item \textbf{The Image Set ($C$):} If a network is driven by a fluctuating environment, it may eventually fall into a restricted subset of states from which certain input sequences cannot easily escape. In phase-space topology, this is an \textit{attractor basin}. In KR theory, it is an image set.
    \item \textbf{The Permutator Group:} Suppose the network oscillates perfectly between three exact micro-states (e.g., State 1 $\to$ State 2 $\to$ State 3). The mathematical group that describes the reversible shuffling of these exact, raw physical states is the permutator group.
    \item \textbf{Tiles and the Holonomy Group:} Biological systems are rarely perfectly precise; they oscillate between regions of phase space rather than exact points. SgpDec partitions the image set into \textit{tiles}—coarse-grained macrostates within the attractor basin. The group that describes the shuffling of these macroscopic tiles, ignoring the microscopic noise inside them, is the \textit{holonomy group}. 
\end{itemize}
When the permutator and holonomy groups match, the circuit's cycle is perfectly precise. When they differ, the circuit possesses a macroscopic oscillation with microscopic slop.

\subsection{Transversal Representatives (Hidden Symmetries)}
Algorithms designed to analyze monoids often condense their outputs to save memory, listing group components only at ``chosen transversal representatives.'' 

In physical terms, this reflects spatial symmetry. If a neural network exhibits a robust $Z_2$ (period-two) oscillation in one corner of its phase space under a specific environmental drive, it often possesses a physically identical, symmetrical oscillation mirrored in a different corner of the phase space under a different drive. Mathematically, these are \textit{conjugate} (isomorphic along a subduction class). One must manually trace these subduction classes to fully map the neural circuit's symmetrical behaviors across its entire state space.

\subsection{The Round-Trip Checksum}
To guarantee that the algebraic decomposition has not distorted the biological circuit, we utilize a mathematical checksum:
% $$ \mathrm{AsHolonomyTransformation}\bigl(\mathrm{AsHolonomyCascade}(t,\mathrm{sk}),\,\mathrm{sk}\bigr)=t $$
\begin{align*}
&\mathrm{AsHolonomyTransformation}\bigl(
\nonumber\\
&\qquad\mathrm{AsHolonomyCascade}(t,\mathrm{sk}),\,\mathrm{sk}\bigr)=t,
\end{align*}
Reading this nested function from the inside out reveals its physical purpose:
\begin{enumerate}
    \item \textbf{Deconstruction:} We take a raw, physical state transition ($t$) and pass it through $\mathrm{AsHolonomyCascade}$, which uses the system's structural blueprint (the skeleton, $\mathrm{sk}$) to break the physical transition down into its abstract KR ``prime factors.''
    \item \textbf{Reconstruction:} We take those abstract coordinates and pass them through $\mathrm{AsHolonomyTransformation}$, forcing the software to reassemble the physical transition from the abstract hierarchy.
    \item \textbf{Verification:} Perfect equivalence ($=t$) proves that no dynamical information was lost. The algebraic cascade is a flawless simulation of the biological network.
\end{enumerate}

\subsection{Anatomy of a Combinatorial Explosion}
For isolated motifs or simple feedforward cascades (DN-to-WTA), this decomposition is computationally lightweight. However, for the WTA-to-DN cascade, the coordinate state space exploded to $5.78\times10^{21}$ tuples, exhausting available machine memory. 

This is not a failure of the algorithm, but a profound proof of the system's complexity. The WTA-to-DN circuit is a highly recurrent, two-way biological feedback loop (normalization modulates competition, and the winning competitor actively reshapes normalization). The Krohn--Rhodes theorem, however, mandates that \textit{all} systems be represented as a strict, one-way, top-down cascade. 

To represent a recurrent feedback loop using only a feedforward hierarchy, the algorithm must ``unroll'' the loop, stacking redundant layer upon redundant layer to maintain the system's temporal memory. Flattening a tangled biological recurrence into a perfectly ordered algebraic hierarchy requires an astronomical coordinate expansion. This $10^{21}$ combinatorial explosion rigorously underscores how deeply and inextricably entangled the normalization and selection motifs become when forced to interact dynamically.

\appendix

\section{Explicit generators and state encodings}
\label{app:generators}

\subsection{Divisive normalization}

The three-state baseline model has \(Q=\{0,1,2\}\) and update
\(s'=\min(\lfloor 2d/(1+s)\rfloor,2)\), giving generators
\[
f_0=(0,0,0),
\qquad
f_1=(2,1,0),
\qquad
f_2=(2,2,1).
\]
To read these tuples, the position in the list represents the \textit{current} state, and the number at that position represents the \textit{future} state. For example, in $f_1$, a current state of $0$ maps to $2$, while a current state of $2$ maps to $0$. Biologically, under an intermediate drive ($d=1$), a low firing rate jumps to a high rate, and a high rate is heavily shunted back to a low rate.

The intermediate-drive map satisfies \(f_1^2=\mathrm{id}\), so the order-two
component is explicit in a generator; the generated monoid has \(|M_D^{(3)}|=13\)
elements with group of units
\(U=\{(0,1,2),(2,1,0)\}\cong\mathbb Z_2\). Because applying $f_1$ twice returns the system exactly to its starting point, this isolated biological rule already constitutes a perfect, reversible period-two oscillator. It is impossible to test for ``emergent'' composite cycles if the foundational building block inherently oscillates.

The four-state model used throughout the composite analyses has
\(X_D=\{0,1,2,3\}\) and
\(s'=Q_{\mathrm{round}}(2d/(2+s))\) clipped to \(X_D\), giving
\[
f^D_0=(0,0,0,0),
\quad
f^D_1=(1,1,0,0),
\]
\[
f^D_2=(2,1,1,1),
\quad
f^D_3=(3,2,2,1).
\]
Notice that in $f^D_1$, every state maps into either $0$ or $1$, trapping the system. If you apply the rule again, it remains trapped. All four are aperiodic. This mathematically guarantees that the isolated normalization generators are strictly dissipative—acting exclusively as sinks. The generated monoid has \(|M_D|=24\) elements, \(8\)
idempotents, trivial group of units, and \(3\) non-aperiodic elements. The
shortest witness word is \(D1\to D3\to D3\), which generates the local
two-cycle \(2\leftrightarrow 1\). Thus, an oscillation can only be sustained by rapidly fluctuating the external environmental drive.

\subsection{Winner-take-all}

WTA states \((x_1,x_2,y)\in\{0,1\}^3\) are encoded as \(q=4x_1+2x_2+y\):
\[
0=(0,0,0),\quad 1=(0,0,1),\quad 2=(0,1,0),\quad 3=(0,1,1),
\]
\[
4=(1,0,0),\quad 5=(1,0,1),\quad 6=(1,1,0),\quad 7=(1,1,1).
\]
This encoding acts as a simple binary-to-decimal converter mapping the physical geometry of the circuit. The variables $x_1$ and $x_2$ represent the two competing excitatory populations (the ``fours'' and ``twos'' place), while $y$ represents the shared inhibitory interneuron (the ``ones'' place).

With the parameters of the Methods
(\(G=1\), \(S=L=B=\theta_E=\tfrac12\), \(W=\tfrac12\), \(R=0\),
\(\theta_I=\tfrac12\)), the input-conditioned generators are
\begin{align}
f^W_{00}&=(0,0,3,1,5,1,1,1),
\nonumber\\
f^W_{10}&=(4,4,7,1,5,5,5,5),
\nonumber\\
f^W_{01}&=(2,2,3,3,7,1,3,3),
\nonumber\\
f^W_{11}&=(6,6,7,3,7,5,7,7).
\nonumber
\end{align}
Each is aperiodic. The shortest non-aperiodic witness is
\(h=f^W_{10}\circ f^W_{00}=(4,4,1,4,5,4,4,4)\), whose square
\(h^2=(5,5,4,5,4,5,5,5)\) is idempotent with
\(\operatorname{im}(h^2)=\{4,5\}\); on that image \(h\) acts as the
transposition \((4\;5)\), so \(\langle h|_{\{4,5\}}\rangle\cong\mathbb Z_2\). 

Physically, the system collapses into an attractor basin consisting entirely of State 4 ($1,0,0$) and State 5 ($1,0,1$). Looking at the binary decoding, the active competitor ($x_1=1$) never changes. The oscillation is driven entirely by $y$ flipping between $0$ and $1$. This proves the isolated WTA circuit possesses a local rhythm driven strictly by the firing and resting of the inhibitory interneuron.

\subsection{Composite state space}

Composite states \((d,w)\in X_D\times X_W\) are encoded as
\(\operatorname{idx}(d,w)=8d+w\), with inverse
\(d=\lfloor \operatorname{idx}/8\rfloor\), \(w=\operatorname{idx}\bmod 8\).
The two states of the key composite cycle are therefore
\[
4=(D{:}0,\,W{:}4)=(D{:}0,\,(1,0,0)),
\]
\[
13=(D{:}1,\,W{:}5)=(D{:}1,\,(1,0,1)).
\]

Decoding these states reveals the precise physical anatomy of the emergent cycle. In State 4, normalization is minimal ($D=0$), excitatory population 1 is firing, and inhibition is silent, resulting in a state of high gain and unchecked excitation. In State 13, normalization is elevated ($D=1$), excitation remains active, and the inhibitory gate fires. Because both the normalization context ($D$) and the competitive identity ($W$) change simultaneously, this cycle represents a genuinely composite, rhythmic ``breathing'' pattern: excitation triggers the inhibitory gate, which simultaneously raises the normalization penalty; this combined suppression then resets the gate, dropping the normalization penalty, allowing excitation to surge again.

Note that \(f^D_0\) is the constant map to \(0\), so the independent product
and the DN-to-WTA cascade share their \(\gamma=0\) generator; the two systems
differ only through the generators with \(\gamma>0\).

\section{Rank distributions}
\label{app:ranks}
Table~\ref{tab:ranks} gives the rank-stratified element counts. The rank of a
transformation is \(\operatorname{rank}(f)=|\operatorname{Im}(f)|\), where
\begin{itemize}\itemsep0pt
    \item $f$: A specific transformation (a sequence of stimuli),

    \item $\operatorname{Im}(f)$: The "Image" of the transformation. This is the set of all possible states the system can end up in after the transformation is applied.

    \item $\vert{} \dots \vert{}$: The mathematical cardinality (how many items are in the set).
\end{itemize}

In physical terms, rank measures the ``width'' of the surviving phase space after a transformation is applied, telling us how much information the system has dissipated. If you start with a 32-state composite system and apply a transformation, how many possible states are left? $Rank 32$ means the system is perfectly reversible. No states collided. (Hamiltonian/conservative). $Rank 2$ shows that the system aggressively funneled 32 possible initial conditions down into just 2 surviving states. (Highly dissipative), whereas $Rank 1$ indicates total collapse. No matter where the system started, it ended up in one single, inescapable fixed point.

In every system with non-aperiodic elements, those elements are concentrated at low
rank: local reversible structure appears only on images reached after
irreversible collapse. Dynamically, this means the biological circuits do not act as global, perfectly conservative oscillators. Instead, they act as funnels: they aggressively dissipate transient states and noise, compressing the full phase space down into a narrow attractor basin. Only once the system is trapped at the bottom of this low-rank valley does it express periodic, rhythmic cycles. This is why the global group of units (transformations that are reversible across the entire, uncollapsed phase space) is trivial in all systems except the three-state baseline.

\begin{table}[t]
\centering
\begin{ruledtabular}
\begin{tabular}{llr}
System & Rank & Elements (non-aperiodic) \\
\colrule
Four-state DN
 & 1 & 4 (0) \\
 & 2 & 18 (3) \\
 & 3 & 1 (0) \\
 & 4 & 1 (0) \\
\colrule
WTA
 & 1 & 8 (0) \\
 & 2 & 272 (43) \\
 & 3 & 35 (4) \\
 & 4 & 10 (2) \\
 & 8 & 1 (0) \\
\colrule
Independent product
 & 1 & 22 (0) \\
 & 2 & 490 (77) \\
 & 3 & 32 (4) \\
 & 4 & 17 (5) \\
 & 6 & 5 (0) \\
 & 8 & 8 (2) \\
 & 12 & 1 (0) \\
 & 32 & 1 (0) \\
\colrule
DN-to-WTA cascade
 & 1 & 15 (0) \\
 & 2 & 2124 (237) \\
 & 3 & 887 (134) \\
 & 4 & 113 (27) \\
 & 5 & 4 (3) \\
 & 6 & 1 (0) \\
 & 7 & 1 (0) \\
 & 8 & 2 (0) \\
 & 12 & 1 (0) \\
 & 32 & 1 (0) \\
\colrule
WTA-to-DN cascade
 & 1 & 17 (0) \\
 & 2 & 1548 (159) \\
 & 3 & 1427 (179) \\
 & 4 & 72 (17) \\
 & 5 & 6 (1) \\
 & 6 & 5 (3) \\
 & 7 & 4 (2) \\
 & 10 & 1 (0) \\
 & 11 & 2 (0) \\
 & 12 & 1 (0) \\
 & 32 & 1 (0) \\
\colrule
Synchronous recurrent
 & 5 & 3 (2) \\
 & 6 & 1 (1) \\
 & 7 & 1 (1) \\
 & 8 & 1 (0) \\
 & 11 & 1 (1) \\
 & 16 & 1 (1) \\
 & 32 & 1 (0) \\
\end{tabular}
\end{ruledtabular}

\caption{
Rank distribution of each generated transition monoid, with the number of
non-aperiodic elements at each rank in parentheses. The cascades heavily concentrate their cycles at Rank 2 and 3, reflecting deep attractor basins. The asynchronous recurrent monoids are omitted: each has eight elements and no non-aperiodic elements. This total loss of cyclic structure upon breaking temporal symmetry proves that the high-rank oscillations seen in the synchronous recurrent model are mathematical artifacts of instantaneous updating, rather than robust physical properties.
}
\label{tab:ranks}
\end{table}

The WTA-to-DN cascade Table.\ref{tab:ranks} reads:
\begin{itemize}\itemsep0pt
    \item Rank 32: $1 (0)$. There is exactly 1 transformation that preserves all 32 states (the identity). It is not an oscillation ($0$).
    \item Rank 12 down to Rank 4: You see elements here, but very few are cycles. The system is still contracting.
    \item Rank 2 $\&$ 3: $1548 (159)$ and $1427 (179)$. This is the attractor floor. The vast majority of the computational transformations map the system down to exactly 2 or 3 surviving states. It is inside these deep, low-rank valleys that the 159 + 179 non-aperiodic cycles emerge.
    \item Rank 1: $17 (0)$. There are 17 sequences that completely kill the system, driving it into a single, inescapable fixed point. By definition, a single fixed point cannot oscillate, so the parentheses are always $(0)$ here.
\end{itemize}

\end{document}